\documentclass[showpacs,twocolumn,pre,eqsecnum]{revtex4}

\usepackage{amsmath}
\usepackage{amssymb}
\usepackage[T1]{fontenc}
\usepackage{dcolumn}
\usepackage{graphicx}

\def\barmu{\bar \mu}
\def\barp{\bar p}
\def\bphi{\bar \phi}

\def\s{\sigma}
\newcommand{\I}{\mathcal{I}}

\newfont{\boldit}{cmbxti10 scaled \magstep 5}
\newfont{\bolditt}{cmbxti10 scaled \magstep 2}
\newcommand{\barf}{\bar{f}}
\newcommand{\HC}{{\scriptscriptstyle{\rm{HC}}}}
\newcommand{\nDH}{{\scriptscriptstyle{\rm{DH}}}}
\newcommand{\Id}{{\scriptscriptstyle{\rm{Id}}}}
\newcommand{\nFL}{{\scriptscriptstyle{\rm{FL}}}}
\newcommand{\MSA}{{\scriptscriptstyle{\rm{MSA}}}}
\newcommand{\Bj}{{\scriptscriptstyle{\rm{Bj}}}}
\newcommand{\bI}{\textbf{I}}
\newcommand{\vecr}{\mathbf{r}}
\newcommand{\Te}{{T^*}}
\newcommand{\Tce}{{T^*_c}}
\newcommand{\cO}{\mathcal{O}}
\newcommand{\lm}{{l,m}}
\newcommand{\Ylm}{Y_{\lm}}
\newcommand{\Alm}{A_{\lm}}
\newcommand{\Qlm}{Q_{\lm}}
\newcommand{\Qlms}{Q_{\lm}^{\sigma}}
\newcommand{\El}{{\scriptscriptstyle{\rm{El}}}}
\newcommand{\pe}{\!=\!}
\newcommand{\as}{a_{\sigma}}
\newcommand{\roc}{{\rho_c}}
\newcommand{\roce}{{\rho_c^*}}
\newcommand{\xid}{\xi_{{\scriptscriptstyle{\rm{D}}}}}
\newcommand{\kb}{k_{{\scriptscriptstyle{\rm{B}}}}}

\newcommand{\chik}{\chi^{(k)}}
\newcommand{\Tr}{(T,\rho)}
\newcommand{\rok}{\rho^{(k)}}
\newcommand{\Tak}{T_{a, k}}
\newcommand{\roak}{\rho_{a, k}}

\begin{document}

\title{Criticality in Charge-asymmetric Hard-sphere Ionic Fluids}

\author{Jean-No\"el Aqua}

\altaffiliation[Now at ]{\'Ecole normale supérieure de Lyon, 69364 Lyon, \textsc{France}}
\author{Shubho Banerjee}

\altaffiliation[Now at ]{Department of Physics, Rhodes College, Memphis,
  Tenessee 38112}
\author{Michael E. Fisher}

\affiliation{Institute for Physical Science and Technology,\\
University of Maryland, College Park MD 20742 (USA)}

\date{\today}

\begin{abstract}

Phase separation and criticality are analyzed in $z$:1 charge-asymmetric
ionic fluids of equisized hard spheres by  generalizing the
Debye-H\"{u}ckel approach combined with ionic association, cluster
solvation by charged ions, and hard-core interactions, following lines
developed by Fisher and Levin
(1993, 1996) for the 1:1 case (i.e., the restricted primitive
model).  Explicit analytical calculations for 2:1 and
3:1 systems account for ionic association into dimers, trimers, and
tetramers and subsequent multipolar cluster solvation. The reduced
critical
temperatures, $T_c^*$ (normalized by $z$), \textit{decrease} with charge
asymmetry, while the critical densities \textit{increase} rapidly with
$z$. The results
compare favorably with simulations and represent a distinct
improvement over all current theories such as the MSA, SPB, etc.  For
$z$$\ne$$1$, the interphase Galvani (or absolute electrostatic) potential
difference, $\Delta \phi(T)$, between coexisting liquid and vapor
phases is calculated and found to vanish as $|T-T_c|^\beta$ when
$T\rightarrow T_c-$ with, since our approximations are classical,
$\beta\pe\frac{1}{2}$. Above $T_c$, the compressibility maxima
and so-called $k$-inflection loci
(which aid the fast and accurate determination of the critical
parameters) are found to exhibit a strong $z$-dependence.

\end{abstract}


\maketitle

\section{Introduction}
\label{intro}

The location and nature of criticality in ionic fluids have been
subjects of intense interest in recent
years \cite{mef1,stell1,ws}. At sufficiently low temperatures
fluid electrolytes typically undergo separation into low and high
concentration phases which may be driven primarily by the
Coulombic interactions. The universality class of the associated
critical points has been under debate owing to apparently
conflicting experiments, inconclusive simulations, and the
analytic intractability of the statistical mechanics beyond a mean
field level \cite{mef1,stell1,ws}. Possible scenarios that have
been discussed include, classical or van der Waals critical
behavior (as might be anticipated in view of the long-range
Coulomb forces), crossover from classical to Ising-type behavior
sufficiently close to the critical point \cite{ws,gutk&anis01} and, as
the leading candidate, three-dimensional Ising-type criticality
(as might be expected for effective short range interactions
arising from Debye screening): indeed, recent
simulations \cite{luij&mef02,young&mef&luij03,young&mef04}
definitively establish Ising behavior for
the simplest charge and size-symmetric model, namely, the
restricted primitive model (or RPM); but for $z$:1 and size-nonsymmetric
systems, the issue is not yet settled.

The most basic continuum models of ionic fluids are the so-called
two-component primitive models consisting of $N \pe N_+ $+$ N_-$
hard spheres, $N_+$ carrying a charge $q_+ \pe z_+q_0$ and $N_-$ a
charge $q_- \pe -z_-q_0$ (with $N_-/N_+ \pe z_+/z_- \equiv z$ so that the
overall
system is electrically neutral). The background medium is assigned
a uniform dielectric constant, $D$, that may be used to represent
a nonionic solvent. In the simple cases on which we focus here,
all the spheres have the same diameter, i.e.
$a_{++} \pe a_{+-} \pe a_{--} \pe a$. The natural and most appropriate
reduced
temperature variable is then determined by the contact energy of a $+z q_0$
ion with a $-q_0$ counter-ion so that
\begin{equation}
\label{T*}
T^*\equiv {\kb T D a_{+-}/ q_+ |q_-|}= {\kb T D a / z q_0^2}\,.
\end{equation}
Likewise, the
normalized density is reasonably taken as
\begin{equation}
       \label{rho*}
       \rho^* \equiv N {a_{+ -}}^{\! \! \! \! \! \! \! \!  \! 3} \quad   \! / V =
{\rho a^3 } \, ,
\end{equation}
in which $V$ is the total volume.

This model (with many ionic species) was first analyzed by Debye
and H\"{u}ckel (DH) \cite{deby&huck23} who derived an approximate
expression for the Helmholtz free energy by solving the linearized
Poisson-Boltzmann equation for the potential around each hard-core ion. For
the simplest 1:1 (or $z \pe 1$) case, i.e.,  the \textit{restricted
primitive model} (RPM), the DH theory predicts \cite{mef&levi93,levi&mef96}
a critical
temperature, $T_{c,{\nDH}}^* \pe \frac{1}{16} \pe 0.0625$, that is in
surprisingly good agreement with modern simulations
\cite{luij&mef02,young&mef&luij03,young&mef04,cail&leve97,orko&pana99,yan&pabl99}
that yield $T_c^*\lesssim 0.05$; however, the critical density predicted by
the
DH theory, namely,
$\rho_{c,{\nDH}}^*\pe{1/ 64\pi}\simeq 0.005$, is significantly too low 
since the
simulations indicate $\rho_{c}^*\gtrsim 0.07$.  Because ionic criticality 
occurs at such low temperatures, the association of charges of opposite signs
into `clusters' is an essential feature in the strongly interacting regime,
as observed in criticality and phase separation in other Coulomb
systems \cite{lika01}. 
Hence, the first crucial improvement  contributed by 
Fisher and Levin (FL) \cite{mef&levi93,levi&mef96}
was to incorporate Bjerrum ion pairing \cite{bjer26}
into the DH theory: this then depletes the density of the free ions that
drive the transition, as a result of which the predicted critical density
increases by a factor of $9$. However, in order to get an acceptable phase 
diagram, Fisher and Levin also found it essential to
account for the \textit{solvation} of the dipolar ion pairs, or
\textit{dimers}, by the residual ionic fluid. The resulting
``DHBjDI'' theory (with `DI' signifying dipole-ionic-fluid solvation)
yields critical parameters, namely, $T_{c,{\nFL}}^*\simeq 0.055$-$0.057$, 
$\rho_{c,{\nFL}}^*\simeq 0.026$-$0.028$ which, to date,
provide the best agreement with the simulations (which indicate 
$\Tce \simeq 0.0493_3$, $\roce \simeq 0.075$
\cite{young&mef04}).

The other most commonly used
theory, the mean spherical approximation
(MSA) \cite{wais&lebo72,telo&evan80,sabi&bhui98,gonz99}, yields
$T_{c,{\MSA}}^*\pe 0.0785$
(even higher than the simple DH theory) and $\rho_{c,{\MSA}}^*\pe
0.0145$ (via the energy route). Although, like the other
approximate theories, the FL approach makes no reliable statements
regarding the universality class of the criticality --- only
classical behavior arises \cite{mef&lee96} --- it does provide
significant physical insights into the origin and location of the
critical point, specifically identifying the role of ionic
association, of the solvation of neutral clusters, and of excluded-volume
effects.

Two generalizations of the RPM are of profound interest, namely,
the \textit{size-asymmetric} primitive model (or SAPM) and the {\em
charge-asymmetric} primitive model (or CAPM). Indeed, it has been
argued \cite{stell1}, that destroying the (artificial) size symmetry of the
RPM might even affect the universality class of the criticality. It may
be suspected that  this feature will eventually be
ruled out by precise simulations. Nevertheless, it has
been demonstrated via exactly soluble ionic spherical 
models \cite{jn&mef04PRL}, that size-asymmetry
can produce dramatic changes: explicitly, the charge correlations become
``infected'' by the critical density fluctuations leading to 
the destruction of normal Debye exponential screening
\textit{at} criticality. Hence, asymmetry should be
carefully accounted for in any realistic analyses of ionic criticality.

In the size-asymmetric model the $+$ and $-$ ions have unequal
diameters: computer simulations \cite{rome&orko00,yan&pabl01} then indicate
that both $T_c^*$ and $\rho_c^*$ \textit{decrease} with increasing size
asymmetry. However, this is directly \textit{opposite} to the trends
predicted by the MSA and some of its
extensions \cite{gonz99,rain&rout00}.  On the other hand, a DH-based
theory developed by Zuckerman, Fisher and Bekiranov \cite{zuck&mef01}
that recognizes the crucial existence of ``border zones'' around each
ion in which the charge is necessarily unbalanced, does, in fact, predict
the correct initial trends, as does
the ionic spherical model \cite{jn&mef04PRL}.

Here we study \textit{charge-asymmetric models} in which the
diameters of the basic positive and negative ions remain \textit{equal}, but
the charges are in the ratio $z$$:$$1$ ($z_- \pe 1$).
Although this model is somewhat artificial for applications to,
for example, multivalent molten salts such as CaF$_2$, or AlCl$_3$
(since, in actuality, the cation and anion sizes are rarely
equal), it nonetheless, represents a valuable step in searching for
a physical understanding of real systems \cite{levi02} 
which exhibit both charge
and size asymmetry (as well, of course, as other complexities such
as short range attractions, etc.).

One should remark, first, that
with the normalizations \eqref{T*} and \eqref{rho*}, the original
DH theory \cite{deby&huck23} predicts that $T_c^*(z)$ and $\rho_c^*(z)$ are
\textit{independent} of $z$; furthermore, the same is true for the MSA 
\cite{sabi&bhui98,gonz99}. However, we
attack the problem via an approach which extends the DH-based
methods developed by Fisher and Levin \cite{mef&levi93,levi&mef96} 
for the RPM as
sketched above. Specifically, we calculate approximate
critical parameters and
coexistence curves for 2:1 and 3:1 systems  by explicitly
accounting for the association of the individual ions into
dimeric, trimeric, and tetrameric neutral \textit{and} charged clusters,
and by including the multipolar cluster solvation free energies
induced by the ionic medium. In the calculations reported here  
the excluded volume effects associated with the hard-core ion-ion repulsion 
enter
in three crucial ways: first, in the solvation free energies of the
individual
ions, as in the original DH theory, and of the neutral and charged ionic
clusters, as in FL; secondly, in the
computation of the cluster association constants which play a pivotal role;
finally, via general hard-core `virial terms' in the free energy 
(described within a simple free-volume approximation 
\cite{levi&mef96}).

\begin{figure}[t]
\scalebox{0.43}{\includegraphics{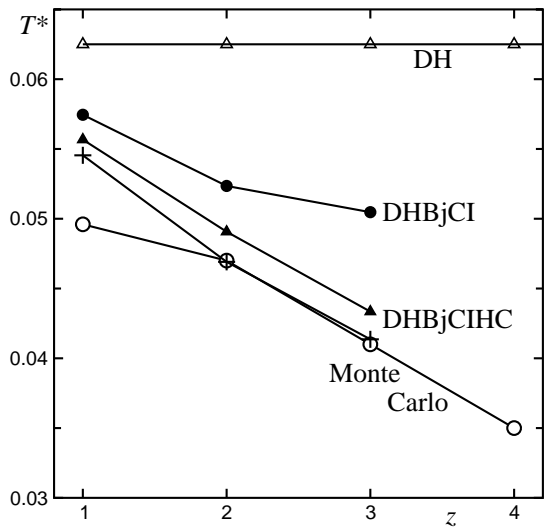}}
  \caption{\label{fig-Tc2z} Critical temperatures as a function of charge
    asymmetry, $z$, as predicted by the present DHBjCI theory [see Eqs.
    \eqref{su}, \eqref{sq} and \eqref{ss}] and its
    refinements including hard-core (HC) virial terms with `standard'
    (triangles: see Table \ref{table}) and `optimal fit' parameters [crosses: see Eqs.
    \eqref{sd}, \eqref{sc} and \eqref{sn}],
    compared with Monte Carlo simulations
    \cite{camp&pate99,pana&mef02} (open circles) and the original
    Debye-H\"uckel (DH) theory.
    The specific parameter values entering the calculations are discussed in
    Sec.~\ref{results}: see \eqref{sd}, \eqref{sc}, etc.}
\end{figure}

\begin{figure}[t]
\scalebox{0.43}{\includegraphics{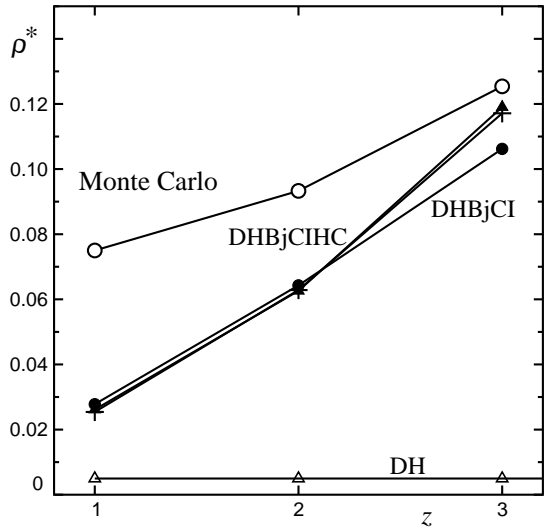}}
  \caption{\label{fig-rc2z} Critical densities as a function of
    charge asymmetry $z$, as predicted by the DHBjCI theory and its
    refinements, using the same conventions as in Fig.~\ref{fig-Tc2z}.}
\end{figure}

In its primary version our theory may be dubbed a DHBjCI approach, with
`CI' signifying cluster-ionic-fluid solvation including the neutral
($z \!+\!1$)-mer and all smaller charged clusters. When specific hard-core 
(HC) excluded-volume virial terms are included, we will label the theory 
DHBjCIHC. More detailed specific refinements will also be examined in order 
to understand the interplay of various effects. However, in all versions,
our approach unambiguously predicts that the \textit{critical temperatures}, 
$\Tce (z)$, \textit{decrease} with increasing charge asymmetry, $z$, while 
the critical densities, $\roce (z)$, \textit{increase} markedly. This
behavior is exhibited in Figs.~\ref{fig-Tc2z} and \ref{fig-rc2z} and 
clearly contrasts with the $z$-independence predicted by the DH and MSA
approximations. Furthermore, one sees from the figures that our results 
mirror
closely the trends uncovered by simulations \cite{camp&pate99,pana&mef02}. 

The main physical effect behind
these trends appears to be that increasing the charge asymmetry
produces a larger number of neutral and charged, but relatively
inert ion clusters: the depletion of the density of (charged) ions
and their smaller average mean-square charge leads, first, to a lower
critical temperature, and, thereby, as in the 1:1 case, to a
higher critical density. In Sec.~\ref{discussion} we explore
this interpretation further and present a comparison with other
current theories \cite{sabi&bhui98,netz&orla99,cail05}:
these either fail to yield even the correct sign of the changes with $z$ or
else predict effects that are much too small!

In order to obtain efficiently, accurate numerical values 
for the critical parameters implied by the DHBjCI theories, 
we have utilized the so-called
$k$-inflection loci introduced recently \cite{orko&mef01,young&mef&orko03}. 
These are defined as the loci on which
$\chi^{(k)} = \chi \Tr / \rho^k$ is maximal  at fixed $T$ above $T_c$, where
$\chi (T, \rho) \! = \!\rho (\partial \rho/ \partial p )_T$.
These loci all intersect at the critical
point but their behavior is also of interest on a larger scale: See 
Figs.~\ref{fig-chik-DH} and 
\ref{fig-rmk} below. In our analysis we find that they 
are strongly dependent on the details of the model (such as 
the hard cores) as well as on the charge asymmetry. 
Thus for our preferred parameters, the values of $k$ for which 
the $k$-locus has  a vertical slope at $(T_c, \roc)$ are 
$k_0 (z)\! \simeq \!0.93$, 
$0.18$, and $-0.87$ for $z \pe 1$, $2$, and $3$, respectively. It should 
be possible to check these predicted trends via simulations.  

An interesting new feature that arises in our calculations (but is
absent by symmetry in the RPM) is the appearance of a nontrivial
electrostatic potential difference, $\Delta \phi(T)$, between
coexisting liquid and vapor phases when $z \ne 1$. This
electrostatic potential, appropriately deemed a \textit{Galvani
potential} \cite{sparbock,iosi00}, has been explicitly
anticipated in the case of 1:1 electrolytes with nonsymmetric
ion-ion interactions in an interesting phenomenological
treatment by Muratov \cite{mura01} (that, however, fails to satisfy
the important Stillinger-Lovett sum rule \cite{stil&love68}).
It also features in a detailed discussion
of colloidal systems (with $z\gg 1$) by Warren \cite{warr00}.

However, the dependence of $\Delta \phi$ on $z$ for moderate charge
asymmetry has not been examined previously. On approach to the
critical point we find that $\Delta \phi(T)$ vanishes as
$|T-T_c|^\beta$, where, because our approximations are classical
in character, we obtain $\beta \pe \frac{1}{2}$. (A similar conclusion
is reached for the asymmetric 1:1 electrolyte
in \cite{mura01}). As a consequence of this potential difference,
a charged double layer \cite{sparbock,iosi00} will exist at a two-phase,
liquid-vapor interface; this, in turn, will be of significance for
interfacial properties such as the surface tension, which have
been studied theoretically, to our knowledge, only for the
RPM \cite{weis&schr98,levi00}.

The balance of this article is laid out as follows: in the next section 
pertinent thermodynamic principles are summarized briefly. Sec.~\ref{assoc} 
then describes
the computation of association constants for the ``primary'' set of 
associated clusters
consisting of one cation of charge $+zq_0$ and $m\le z$ anions of charge 
$-q_0$;
detailed calculations for tetramers are presented in Appendix \ref{appa}. The
crucial multipolar electrostatic contributions to the Helmholtz free energy
are analyzed in Sec.~\ref{electrostatic}. These and other ingredients are
combined in Sec.~\ref{para} to obtain expressions for the total free
energy and, thence, in Sec.~\ref{results}
quantitative results for the 2:1 and 3:1
models (following a brief account of the pure DH theory). A discussion
of the $k$-inflection loci is presented in Sec.~\ref{chik}.
Sec.~\ref{galvani} is
devoted to the Galvani potentials while, as mentioned, our results are 
reviewed briefly 
and compared with those of other current theories
in Sec.~\ref{discussion}, the varied predictions for the critical parameters 
being summarized in 
Figs.~\ref{fig-Tc2x} and \ref{fig-rc2x}.




\section{Some Basic Thermodynamics}
\label{main-def}

\subsection{Phase Equilibrium}
\label{phase-eq}

A $z$:1 electrolyte may be regarded
(neglecting the solvent) as a single-component system since
putting $N_0$ neutral `molecules' (each of one positive and $z$
negative ions) at temperature $T$ into a domain of volume $V$
completely defines the thermodynamic state.  The total number of
ions is then $N \pe (z+1)N_0$, while the \textit{total ionic number
density}, $\rho \equiv N/V$, also measures the density of the
original molecules. The total Helmholtz free energy, $F(T,V,N)$
may be introduced, in standard notation, via the differential
relation
\begin{equation}
\label{df} dF = -S dT - p dV + \mu dN\,,
\end{equation}where
$\mu$ is the chemical potential conjugate to the total number of
ions. In the thermodynamic limit, the reduced variables
\begin{multline}
  \barf(T, \rho) \equiv -F/V \kb T,  \quad {\rm  and} \\
  \quad   \barmu(T,\rho) \equiv \mu/\kb T =-(\partial
  \barf/\partial\rho)_T\, ,
\end{multline}
are convenient \cite{mef&levi93,levi&mef96}. The reduced
pressure follows from the variational expression
\begin{equation}
  \label{pressure}
  \barp(T, \mu) \equiv p /\kb T= \max_{\rho}
  \left( \barf(T,\rho)+ \barmu \rho \right)\, .
\end{equation}

Then phase coexistence (if present) at a given temperature is
specified by the equilibrium conditions
\begin{equation}
  \label{equilibrium}
  p(T,\rho_v) = p(T,\rho_l)\quad {\rm and} \quad \mu(T,\rho_v) =
  \mu(T,\rho_l)
  \, ,
\end{equation}
where the subscripts $v$ and $l$ indicate vapor and liquid phases,
respectively. These equations determine the densities in the
two phases: at the critical temperature and density, $\rho_l(T)$ and
$\rho_v(T)$ coincide.

The single-component or molecular thermodynamic formulation takes
care of overall electroneutrality in a natural way and utilizes
only one overall chemical potential. It is complete, in principle,
if one knows the Helmholtz free energy density $\barf(T,\rho)$. An
alternate approach is to treat the ``isolated'' or free ions, and
the various clusters into which they associate, each as distinct
species in thermal equilibrium with one other. Since the exact
calculation of $\barf(T,\rho)$ is intractable, this latter
approach is useful in constructing approximations for the overall
free energy density. Such a formulation, however, requires the
principles of \textit{multi}component thermodynamics that have have
been reviewed systematically by FL for the charge-symmetric RPM
in \cite{levi&mef96} (henceforth abbreviated as {\bI}). The
formulation needed for the charge-asymmetric models is quite
similar to that outlined in {\bI} but contains some subtle
differences. Thus, even at the cost of some repetition, we outline
the main principles here.

Consider a system of distinct species $\s$, which may be free
ions or ion clusters ($\s \pe +,-$ for the original ions, and
$\s \pe 2,3,\cdots$ for dimers, trimers, etc.), with number densities
$\rho_\sigma  \pe N_\sigma/V$, where $N_\sigma$ is the number of
entities of species $\sigma$. The Helmholtz free energy density
$\barf(T;\{\rho_\sigma\})$ can be defined through a generalization
of the single-component formulation above [see Eqs. (2.4) and
(2.5) in {\bI}].  The reduced chemical potential for species
$\sigma$ then follows from 
\begin{equation}
  \label{mu-j}
  \barmu_\sigma(T;\{\rho_\sigma\})\equiv \mu_\s /\kb T = -(\partial
  \barf/\partial\rho_\sigma)\, . 
\end{equation}
Since all the species present will
be in chemical equilibrium, the sum of the chemical potentials of
the reactants in any reaction will equal the sum of the chemical
potentials of the products [see {\bI}(2.2) and {\bI}(2.3)].
These equations together with conditions \eqref{equilibrium} and
overall electroneutrality, namely,
\begin{equation}
  \label{electroneutrality}
  \sum\nolimits_\s q_\s \rho_\s =0\, , 
\end{equation}
determine the system in equilibrium. For calculating the pressure one may 
still use Eq. \eqref{pressure}, or, equivalently, the multicomponent form
{\bI}(2.6).

For a multicomponent system in which none of the species has a net
charge, thermal equilibrium demands that the chemical potentials of
each species match in coexisting phases. More generally,
however, it is the \textit{electrochemical} potential that must be
equal in both phases so that for a species $\s$ one has
\begin{equation}
  \label{eq-galvani} 
   \mu_{\s,v}+ q_\s \phi_v= \mu_{\s,l} + q_\s \phi_l\, ,
\end{equation}
where $q_\s$ is the net charge of particles of species $\s$
and $\phi_v(T)$ and $\phi_l(T)$ denote the (in general distinct)
electrostatic potentials in the coexisting vapor and liquid
phases, respectively. Then $\Delta \phi (T) \equiv \phi_l -
\phi_v$ is the absolute electrostatic potential difference between
the two phases, i.e., the interphase Galvani
potential \cite{sparbock,iosi00}.  In the molecular or `overall'
formalism presented above, the correct phase behavior can be
obtained without any reference to the Galvani potential since the
chemical potential $\mu$, conjugate to the overall density $\rho$,
corresponds to a neutral species that is insensitive to the
electrostatic potential $\phi$. Nevertheless, the Galvani
potential represents a significant feature, that is not present (or 
vanishes identically) in
the RPM: it is discussed in further detail in Sec.~\ref{galvani}.


\subsection {Free Energy Contributions}

Our aim is to construct a physically appropriate, albeit
approximate free energy for the model systems by adding
contributions that arise from the various degrees of freedom and
the underlying mechanisms and interactions. As a zeroth order
approximation any fluid may be taken as an ideal gas. Thus, for
each species we invoke an ideal-gas term 
\begin{equation}
  \label{id-nrg}
  \barf^{\Id}(T;\rho_\s)=\rho_\s -\rho_\s \ln [\rho_\s {\cal
  C}_\s(T)]\, , 
\end{equation}
where $\mathcal{C}_\s(T)$ depends on the internal
configurational partition function of species $\s$, $\zeta_\s(T)$, and the de
Broglie wavelength, $\Lambda_\s(T)$ (see {\bI} for details).

The principal contribution to the interaction free energy of our
model electrolyte comes from the electrostatic interactions
between the ions.  We will use a DH ``charging'' approach to
calculate the electrostatic free energy of each species as discussed in 
detail
in Sec.~\ref{electrostatic} below. The only other significant
interaction between the ions is the hard-core interaction.

The various forms of additive free energy corrections for the hard-core 
contributions that might be employed are discussed at length in
{\bI}. However, we have not explored
the range of these options here. It may be noted, first, that such 
second-order
and higher virial-type corrections \cite{mef&levi93,levi&mef96}, 
enter formally in \textit{higher}
order in powers of the overall density than do the (leading)
electrostatic terms. Secondly, the exact hard-core diameters already
play a quantitatively significant role in DH theory itself (see
Sec.~\ref{pure-DH} below). Furthermore, as  observed in the Introduction,
the exact hard-cores are equally vital in the formation of ion clusters, 
thereby affecting the values of the corresponding association
constants which in turn play a dominant role. Finally, the formation of
tightly bound clusters (see {\bf I} and below) at temperatures
$\lesssim T_c$ has the effect, at the rather low densities near
criticality, of markedly increasing the available free volume
relative to a fluid with only hard-core interactions.

For these reasons, in the present study we have confined our 
considerations to a simple free-volume approximation which adds 
\begin{equation}
  \label{fhc}
  \barf^\HC = ( \sum\nolimits_\tau \rho_\tau ) \ln ( 1 - 
 \sum\nolimits_{\s} B_{\s} \rho_{\s} )\, ,
\end{equation}
to the Helmholtz free energy. In the low density limit with all species
regarded as hard spheres of diameters $\as$, the exact value of the
coefficients $B_\sigma$ is $2 \pi a_\s^3/3$. But, as noted in {\bI}, this 
choice for equisized hard spheres implies an unrealistically low 
maximum density at $\rho^* = 3/2\pi \simeq 0.48$ in contrast to the true, 
fcc packing density of $\rho^*_{{\rm{max}}} = \sqrt{2} \simeq 1.41$. A
reasonable alternative choice for use at intermediate densities is thus to 
take effective values of the $B_\s$ coefficients corresponding to bcc
close packing \cite{mef65}, namely, 
$B_\s / \as^3 = 4/3\sqrt{3} \simeq 0.770$.
We will, hereafter, refer to this choice as using ``bcc hard cores''; its 
influence on the values of the critical parameters will be examined 
below in Sec.~\ref{results}. 

We may remark, however, that while  some improvements in
accounting for volume-exclusion may still be feasible, the studies in
{\bf I} indicate that straightforward, naive approaches tend
to strongly \textit{over-estimate} the excluded volume effects. This seems
to occur because the dominant many-particle ionic correlations in
the low-temperature moderately dense liquid lead to an ``expanded
crystal-like'' structure that screens out direct hard-core
interactions: see plot (d) in Fig.~6 of {\bf I} and the related
discussion in {\bf I} Sec.~8.5.




\section{Association constants for ion clusters}
\label{assoc} 

\begin{figure}[t]
\scalebox{0.45}{\includegraphics{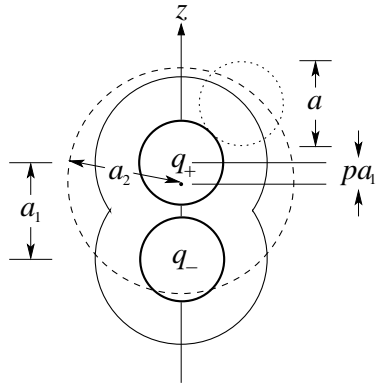}}
  \caption{\label{fig-dimer} A dimer with the two ions separated by a 
    distance $a_1$. The dotted sphere indicates the closest possible
    approach by a screening ion. The dashed sphere of radius $a_2$ 
    represents an effective exclusion zone for solvation computations: 
    Sec.~\ref{galcons}.}
\end{figure}

The success of the DHBjDI theory in estimating the
critical point of the RPM, together with ``snapshots'' of primitive models of
ionic fluids from computer simulations \cite{camp&pate99,pana&mef02} indicate
that a high degree of association is present in the critical
neighborhood. A careful analysis of the association constants for
ion clusters is therefore essential.  Here, we generalize
Bjerrum's original approach \cite{bjer26} to define and calculate
the association constants for a set of ``primary'' clusters which
contain one central cation of charge $z q_0$ surrounded by $1 \leq m
\le z$ singly charged anions (a general dimer configuration is
illustrated
in Fig.~\ref{fig-dimer}). These primary clusters with, including the bare
ions, net charges $q_\s\pe(-1, 0, +1, \cdots, +z)q_0$, will be the
first to form when the temperature is lowered, and the density,
increased. 

Of course other, larger clusters of ions must
eventually come into play. However, insofar as their net charges
fall in the same range, they may be subsumed, in a first
approximation, under the like-charged ions and primary clusters:
see the discussion in {\bI} Secs.~8.3 and 8.4. The most important
exception is probably the \textit{doubly} over-charged ``molecular''
clusters with $m \pe z+2$ counterions: however, we believe that such
clusters will not contribute significantly in the critical region
for $z\le 4$.

\begin{figure}[t]
\scalebox{0.45}{\includegraphics{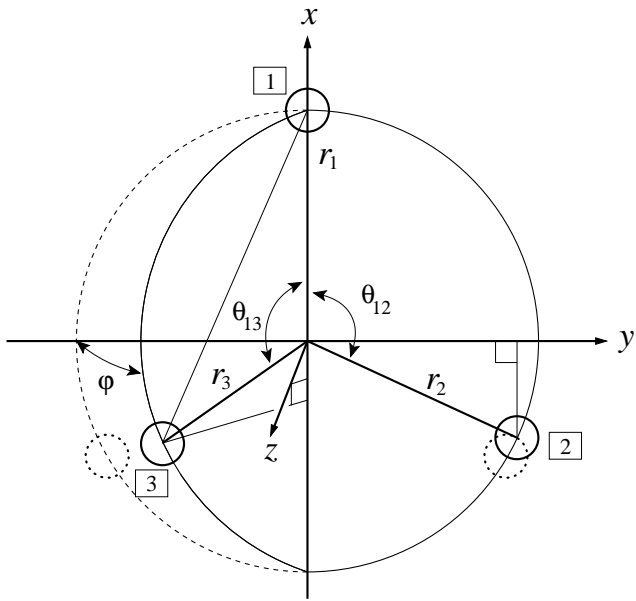}}
  \caption{\label{fig-tetramer} A configuration of a tetramer with
    coordinates suitable for calculating the 
    association constant. The small dotted spheres indicate the 
    ground-state orientations of the satellite ions. The central positive 
    ion is located at the origin.}
\end{figure}

To proceed, consider a charge $q_+ \pe z q_0$ fixed at the origin of
a Cartesian coordinate system with $m$ satellite charges
$q_- \pe -q_0$ around it. For $m \pe 3$ one has a tetramer for which 
Fig.~\ref{fig-tetramer} illustrates a general configuration, and 
let $\vecr_i$ be the position vector of
the $i^{\textrm{-th}}$ 
satellite.  The reduced configurational energy (electrostatic
plus hard-core) for such a system, normalized by $q_+ q_-/ D a$, is
\begin{align}
\label{config-nrg}
 E_{m,z}(\{\vecr_i\}) &= \sum_{i=1}^m \frac{a}{r_i} - 
 \sum_{(i,j)} \frac{a}{z r_{ij}}, \quad \mbox{if} \, \, \, 
 r_i,\, r_{ij} \ge a,  \nonumber \\
   &=  -\infty,  \quad \mbox{otherwise},
\end{align}
where $r_i\pe |\vecr_i|$, $r_{ij}\pe |\vecr_i- \vecr_j|$, 
and $(i,j)$ indicates a sum over all distinct pairs. The
association constant for a cluster or $(m+1)$-mer formed by these
charges is the internal partition function 
\begin{equation}
 \label{assoc-const}
  K_{m,z}(T;R)= \frac{1}{m!} \prod_{i=1}^{m} \int_a^R
  d\vecr_i\, \exp[{E_{m,z}(\{\vecr_i\})/T^*}]\, ,
\end{equation}  
where $R$ is a suitable cut-off radius without which all the integrals would
diverge at large distances.

The choice of $R$ is necessarily somewhat arbitrary since there is
no clear, absolute criterion for when a group of ions is to be
considered ``associated''.  The ambiguity in choosing a cut-off
radius arises even for the simplest possible cluster, a dimer
($m \pe 1$). In that case Bjerrum \cite{bjer26} observed that the
integrand in \eqref{assoc-const} exhibits a minimum at a radius
$R^{\Bj} \pe  a/2T^*$; thus he chose $R \pe R^{\Bj}$ as the cut-off for
$T^*<\frac{1}{2}$ (since for $T^* \ge \frac{1}{2}$ one has 
$R^{\Bj} \le a$). Evidently, the choice $R \pe R^{\Bj}$, makes the
association
constant \textit{least} sensitive to the value of the cut-off.
Bjerrum's choice, however, may reasonably be considered as
unphysical since $R^{\Bj}(T)$ becomes unbounded when $T$ falls to
zero while one expects a $(+,-)$ ion pair to become more tightly
bound at lower temperatures. This issue is discussed in {\bI}
(see Sec.~6.2) where  Bjerrum's association constant is also
compared with other definitions: see also \cite{zuck&mef01,jian&blum02}. While
Bjerrum's cut-off, $R^{\Bj}$, has no direct relevance to the actual
physical size of a dimer which is much more compact (as analyzed
in {\bI}, Sec.~6.2) --- the value and behavior of Bjerrum's
association constant is numerically accurate for $T^*\lesssim \frac{1}{16}$ 
despite the unphysical nature of the cut-off.

In light of this analysis we generalize Bjerrum's approach and choose the
cut-off $R$ so that $(\partial K_{m,z}/ \partial R)$ is minimal.
For dimers,
the choice of the cut-off makes very little difference over a wide
range of $R$ at and below $T^*  \pe \frac{1}{16}$.  However, one may 
anticipate that the dependence of $K_{m = z,z}(T;R)$ for $m\!>\!1$
will be more sensitive to the choice of $R$ because the ground state
binding energy per $(q_+,q_-)$ bond (for a neutral cluster) becomes
smaller with increasing $z$ \cite{erbe&hock97}. As a consequence we must 
expect
our estimates for the critical densities to become less reliable
with increasing charge asymmetry.

It should also be noted that our choice of integration
domain \eqref{assoc-const} is somewhat arbitrary. Thus by taking
the outer boundary surface to be $r_i \pe R$ (for all $i$) we
have chosen to integrate over an $m$-dimensional hypercube.
Instead, one might well choose the $m$-dimensional hypersphere:
$\sum\nolimits_j r_j^2 \le R^2 + (m-1)^2 a^2$ (where $j$ runs from 
$1$ to $m$), or, say, a hypercube cut along its body diagonal,  
$\sum\nolimits_j r_j \le R+(m-1)a$.
However, it is reassuring that for $m \pe 2$, where
exact numerical calculations are possible, the choice of
integration domain makes a difference of less than $0.5\%$ in $K_{2,2}$ at
$T^*_{c,{\nDH}} \pe 0.0625$; furthermore, the sensitivity to this
choice is reduced at lower temperatures.

Now, for small $T$, the integral defining $K_{m,z}$ is dominated
by the ground state energy of the cluster, and, it is
appropriate, therefore, to expand the integrand about the ground state
configuration. After appropriate scaling of the radial variables,
we obtain, for $m \ge 3$, the general form
\begin{multline}
  \label{kmz}
  K_{m,z}(T;R)=\frac{1}{m!} \frac{8\pi^{m+1/2} \mathcal{J}_m a^{3 m}}{
    \prod_{k=1}^{2m-3} \lambda_{m,k}^{1/2}} 
    \times \\
  \frac{z^{m-3/2}\, {T^*}^{2m-3/2}}
  {(C_{m,z})^m} \, \exp(m C_{m,z}/ T^*) \, {\I}_{m,z}(T^*;R)\, ,
\end{multline}
where the residual integral satisfies
\begin{equation}
\label{imz}
{\I}_{m,z}(T^*;R) = 1 + \mathcal{O}(T^*)\, ,
\end{equation}
while the
$\lambda_{m,k}$'s are the eigen\-values of the reduced quadratic form
describing the angular variation of the energy. The dominant
exponential dependence is controlled by $C_{m,z}$, the binding
energy per satellite in units of $(q_+ q_-/ D a)$, while $\mathcal{J}_m$
is the Jacobian of the transformation leading  to the diagonalization
of the angular integrals. In
Appendix \ref{appa} the calculations are performed explicitly for
tetramers ($m \pe 3$: see Fig.~\ref{fig-tetramer})
thereby illustrating the general
procedure. Evidently the principal $T$-dependence of the
association constant can be found by scaling the variables and
expanding the integrand for small $T^*$. A full calculation,
however, requires an evaluation of the residual integral factor
${\I}_{m,z}(\Te;R)$.

Dimers and trimers turn out to be special cases for which the
general form \eqref{kmz} does not apply. Nevertheless, the
calculations follow similar lines.  For dimers the association
constant has been discussed in detail in {\bI}. Expanding around
the ground state yields
\begin{equation}
  \label{k1z}
  K_{1,z}(T;R)= 4 \pi a^3
  T^* \exp[{1/T^*}] \, \I_{1,z}(T^*;R)\, .
\end{equation}
Moreover, using the Bjerrum cut-off $R^{\Bj}$ and evaluating analytically
the integral over $r$ in \eqref{assoc-const} gives \cite{levi&mef96}
\begin{multline}
  \label{i1z}
     \I_{1,z}(T^*;R^{\Bj})= \frac{1}{6 {T^*}^4}
      e^{-1/T^*}[{\rm Ei}(1/T^*)- {\rm Ei}(2)+ e^2]
   \\
     - \frac{1}{6 {T^*}^3} (1+{T^*}+2{T^*}^2)\, ,
\end{multline}
where Ei$(y)$ is the standard
exponential integral.  Because of the normalization (1.1), this
result is independent of $z$.  The asymptotic expansion for small
$T^*$ is, in addition, independent of $R$ and given by
\begin{equation}
  \I_{1,z}(\Te;R)= 1 + 4\,T^* + 4 \cdot 5 \, {T^*}^2 + 4 \cdot 5 \cdot 6
  \, {T^*}^3 + \ldots \, .
\end{equation}
We note that when $T \le 0.1$, this expansion gives reasonably
accurate results if truncated at the smallest term: see
\textbf{I}.

For trimers, a similar but more elaborate calculation yields,
\begin{multline}
  \label{k2z}
  K_{2,z}(\Te;R)= 32 \pi^2 \frac{z {T^*}^3 a^6}{(1-1/4z)^2}
  \\
  \exp[2(1 -1/4z)/T^*] \, \I_{2,z}(\Te;R) \, .
\end{multline}
However, in contrast to
dimers, an exact analytical result for $\I_{2,z}(\Te;R)$ seems
inaccessible for any value of $z$. Nevertheless, one can obtain
precise results by numerical integration. For our subsequent
calculations we need results for trimers with $z\pe 2$ and $3$, i.e., the
integrals $\I_{2,2}$, and $\I_{2,3}$.
Generalizing the Bjerrum procedure, we determine the
appropriate, optimal cutoffs, $R_{m,z}$ by searching numerically
for the minima of $(\partial K_{m,z}/ \partial R)$ at fixed temperature.
The results can be scaled conveniently by setting
$R_{2,z} /a \pe  \tilde{R}_{2,z} (\Te) / T^*$, where, typically, we find
\begin{equation}
  \label{nR2z}
  \tilde{R}_{2,2} (0.05)  \simeq 0.263, \quad
  \tilde{R}_{2,3} (0.05) \simeq 0.336 \, .
\end{equation}
For $z\pe 2$, the sensitivity of the trimer association
constant to the cut-off is markedly greater than found for dimers
(which was illustrated graphically in \textbf{I} Fig.~3).  At a
temperature $T^* \pe  0.052$ (some $6\%$ above the predicted value
of $T_c^{*}(z \pe  2)$: see Sec.~\ref{results}) increasing (decreasing)
$R_{2,2}$ by $20\%$ increases (decreases) $K_{2,2}$ by about $0.7\%$.
However, as is explained in Sec.~\ref{results}, these changes do not
significantly
alter the predicted critical parameters.
Because of the larger central charge, the sensitivity of $I_{2,3}$
to the cut-off is significantly smaller at the relevant temperatures.

Even if one employs a precise numerical calculation of $K_{2,z}$ for
trimers,
it is
worth noting that it can be reproduced quite accurately by Pad\'e
approximants
\cite{mef74} that embody the small-$\Te$ expansion of $\I_{2,z}$. For
example,
when $z \pe 2$, the  expansion
\begin{multline}
  \label{}
    \I_{2,2} (T^*;R) = 1 - \frac{76}{49} \, \Te + \frac{357248}{2401} \,
    \Te^2
  \\
    -\frac{222368768}{117649} \, \Te^3 + \frac{7109382144}{117649} \, \Te^4
    +\ldots \, ,
\end{multline}
yields a $[1/3]$ Pad\'e approximant that up to $\Te \pe 0.055$,
agrees with the results of numerical integration to better than $4\%$. With
this in mind, we consider the approach also for tetramers.

\begin{figure}[t]
\includegraphics[width=8cm]{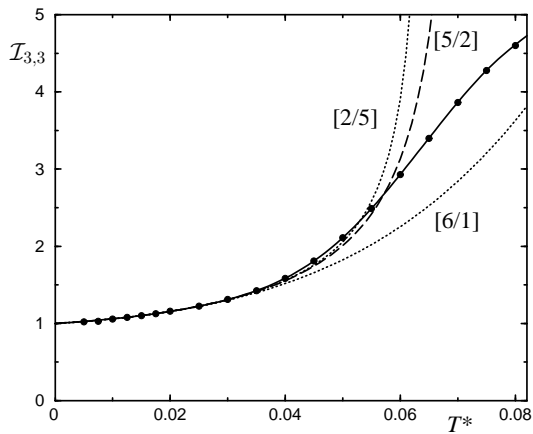}
  \caption{\label{fig-pade} Calculation of the association constant
    integral $\I_{3,3} (T^*)$. The dashed line represents the
    $[5/2]$ Pad\'e approximant while the dotted lines portray the
    $[6/1]$ and $[2/5]$ approximants. The solid circles result from
    Monte Carlo integration while the solid line is a polynomial fit: see
    \eqref{fitItt}.}
\end{figure}

Indeed, for tetramers (needed only for the case $z\pe 3$),
expanding about the ground state yields,
\begin{multline}%
  \label{K33}
    K_{3,z}(\Te;R) = \frac{48}{5} 2^{1/2} 3^{1/4}
    \pi^{7/2} \frac{z^{3/2} a^9 {T^*}^{9/2}}{(1-1/{\sqrt 3} \, z)^3}
    \times  \\
    \exp [(3 - {\sqrt 3 /z})/T^*] \, \I_{3,z}(\Te;R)\, ;
\end{multline}%
See Appendix \ref{appa}. However, the
calculation of ${\I}_{3,z}$ proves difficult even numerically.
Asymptotic expansion for small $T^*$ yields
$\I_{3,3} = \I_{3,3}^{(7)} + \cO (\Te^8)$ with (after some efforts)
\begin{multline}
  \label{seriesItt}
  {\I}_{3,3}^{(7)}(\Te;R)=1 + 4.26324 \, \Te + 157.697 \, \Te^2
  \\
   + 353.407 \, \Te^3   + 29636.117 \, \Te^4 
  - 58642.1 \, \Te^5 \\%
  + 8.5259 .10^6 \, \Te^6 - 7.07815. 10^7 \, \Te^7 \, .
\end{multline}
One can then form and examine all
the approximants up to order $7$. One observes readily that the
$[5/2]$ approximant seems the most reliable judging its convergence
relative to the other approximants: see Fig.~\ref{fig-pade}.

However, since the tetramer is of prime importance for criticality in the
3:1
model, and because one knows that an approximant based only on the low-$T$
asymptotics must fail at some value of $T^*$ (of likely magnitude 
$\sim \! 0.1$), we have undertaken a Monte-Carlo
evaluation of $\I_{3,3}$ 
\cite{goul&tobo}: see Fig.~\ref{fig-pade}. The details are described 
in Appendix \ref{MC}. It transpires that the 
$[5/2]$ Pad\'e 
approximant agrees to within $4\%$ with the precise numerical calculation
up to $\Te \pe  0.06$ (which is $15\%$ higher than the DHBjCI value 
of $\Tce (z \pe 3)$ as can be seen in Fig.~\ref{fig-Tc2z}).
Nevertheless, for our explicit calculations we fitted the Monte Carlo
calculations of $\I_{3,3}$ to the form 
\begin{equation}
  \label{fitItt}
  \I_{3,3} (\Te; R) = \I_{3,3}^{(7)} (\Te; R) + \sum\nolimits_{j=8}^{18}
  i_j \, \Te^j \, ,
\end{equation}
where the coefficients $i_j$ are listed in Table \ref{tableij}. As seen in
Fig.~\ref{fig-pade}, the fit is very good and, indeed, provides an
accuracy of one part in $10^3$ or better. In reality, it probably remains valid some
way above $T^* \pe 0.10$; but it is also clear that the $[5/2]$ approximant
fails rapidly above $T^* \pe 0.06$ and shows noticeable deviations already
for $T^* \! \gtrsim \! 0.04$.

\begin{table}
\caption{\label{tableij} Fitted expansion coefficients $i_j$ for calculating
  the tetramer association constant $K_{3,3} (T^*)$: see \eqref{K33}-
  \eqref{fitItt}.}
\begin{ruledtabular}
\begin{tabular}{lll}
$j \quad \quad ~ 10^{-(j+3)} \, i_j$ & $j \quad \quad ~ 10^{-(j+3)} \, i_j$ & 
  $j \quad~~ 10^{-16}  \, i_j$ \\
$8~~~~~\, -0.419627$ & $12~~\, \quad 55.247$ & $16 \quad  13.8829$ \\
$9~~~~~~ \quad 2.17887$ & $13~~ - 44.2769$ & $17 \quad ~\, 2.58295$ \\
$10~~~~~~~\, 2.6276$ & $14 \quad \,~~ 12.4183$ & $18 \quad ~\, 0.40201$ \\
$11~~\, -28.3178$ & $15 \quad ~~~\, \, 0.558599$ & 
\end{tabular}
\end{ruledtabular}
\end{table}


\section {Electrostatic contributions to the free energy}
\label{electrostatic}

\subsection {General considerations}
\label{galcons}

To calculate the electrostatic part of the free energy we adopt
the basic DH strategy \cite{deby&huck23} but, as in  
\cite{mef&levi93}, we
generalize the approach to include cluster species that contain
more than one ion and, thus, are not spherically symmetric.
Consider a cluster (possibly just a single ion) of species
$\sigma$ that has charges $\{q_i\}$ at positions $\{\bf{r}_i\}$.
Owing to the hard-core repulsions the ``free'' screening ions are
prevented from entering the ``exclusion zone'' of the species: see,
for example, the dimer with one $+2q_0$ and one $-q_0$ ion which
has a dumbbell shaped exclusion zone as seen in see Fig.~\ref{fig-dimer}. 

To
estimate the free energy of an isolated cluster in an atmosphere
of screening ions, of densities $\rho_\nu$ and charges $q_\nu$, we
approximate its exclusion zone by a \textit{sphere} of radius
$a_\sigma$ \cite{mef&levi93,levi&mef96}: for the selection of an 
appropriate value for
$a_\sigma$, see below in Sec.~\ref{geom}. At this point we will suppose
only that the choice of origin for this effective exclusion sphere
is such that all the charges of the cluster are included within
it.  A specific criterion for the precise choice of origin for the
effective exclusion sphere (when not dictated by an
obvious symmetry) will be developed for each cluster species as we
address them individually.

For $r \le a_\sigma$, the overall electrostatic potential may generally 
be expanded in terms of the spherical harmonics $\Ylm$ as
 \begin{multline}%
  \label{inside}
    \Phi_<(r,\theta,\varphi) =
    \frac{1}{D} \sum_{\lm}
    \frac{4\pi}{2l+1}
  \\
    \times \left[ \sum\nolimits_i q_i
    \Ylm^* (\theta_i, \varphi_i) \frac{r_{i,<}^l}{r_{i,>}^{l+1}} \, + \,
    \Alm \, r^l \right] \Ylm (\theta, \varphi)\, ,
\end{multline}%
where $D$ is the dielectric constant of the medium,
$i$ labels the particles of the cluster $\s$, the $q_i$ are their
charges and the
$(r_i,\theta_i,\varphi_i)$, their coordinates, while
$r_{i,<} \pe  \min (r, r_i)$, $r_{i,>} \pe  \max (r, r_i)$, and
the notation $\sum\nolimits_{\lm}$  means
$\sum_{l=0}^\infty \sum_{m=-l}^l$. We note that the
boundary condition at the origin is already taken into account in
this expression.

For $r \geq a_\sigma$ the potential arising
from the cluster ions in $\s$ is screened by the external ions
and hence we may expand the potential as
\begin{equation}
  \label{outside}
  \Phi_> (r, \theta, \varphi)=  \frac{1}{D} \sum_{\lm}
        B_{\lm} \, k_l(\kappa r) \, \Ylm (\theta,\varphi) \, ,
\end{equation}
in which screening is embodied in
the boundary condition $\Phi_> \rightarrow 0$
when $r \rightarrow \infty$ [which relates rather directly 
to the introduction of the electrostatic potential 
$\phi$ in \eqref{eq-galvani}]. The inverse Debye
length introduced here is defined generally by
\begin{equation}
  \label{kappa}
  \kappa(T,\{\rho_\tau\})= \left( 4\pi\sum\nolimits_\tau \rho_\tau q_{\tau}
    ^2/D \kb T \right)^{1/2} \equiv 1/\xid \, ,
\end{equation}
and, when convenient, we will write
\begin{equation}
\kappa a = x \quad \textrm{and}\quad \kappa
a_{\sigma} = x_{\sigma}\, .
\end{equation}
The spherical Bessel functions
\begin{equation}
k_l(x)=g_l(x) e^{-x}/x^{l+1}\, ,
\end{equation}
that arise in the solution
of the Debye-H\"{u}ckel or linearized Poisson-Boltzmann equation,
are conveniently specified in terms of the polynomials
\begin{equation}
g_l(x)=\sum_{m=0}^l \frac{(l+m)!}{2^m \, m! (l-m)!} x^{l-m}\, ,
\end{equation}
(so that $g_0 (x) \pe 1$, $g_1 (x) \pe 1+x$, $g_2(x) \pe 3+3x+x^2$, etc.)

On the surface of the exclusion sphere, $r\pe a_\sigma$, matching
$\Phi_<$ and $\nabla \Phi_<$ to $\Phi_>$ and $\nabla \Phi_>$
(the usual conditions expressing continuity of the potential  and
absence of surface charge)
yields the coefficients
\begin{equation}
  \label{alm}
  \Alm = -\frac{\Qlm}{a_\s^{2l+1}} \left[  1-
  \frac{(2l+1)k_l(x_\s)}{x_\s \, k_{l+1}(x_\s)} \right]  \, ,
\end{equation}
\begin{equation}
  \label{blm}
  B_{\lm}= \frac{4 \pi \, \Qlm}{a_\s^{l+1} \, x_{\s} \, k_{l+1}(x_\s)}\, ,
\end{equation}
in which  the cluster
multipole moments, $\Qlm$, which will play a central role in our
calculations, are given by
\begin{equation}
  \label{qlm}
  \Qlm = \sum\nolimits_{i}
  \Ylm^* (\theta_i,\varphi_i)\, q_i\, {r_i}^l\, ,
\end{equation}
where the summation runs over the particles  of the cluster $\s$.
In \eqref{inside}, the potential arising directly from the ions
in the cluster (without any contribution from the screening ions) is
 \begin{multline}%
  \label{bare}
   \Phi_0 (r,\theta,\varphi) =
   \frac{1}{D} \sum_{\lm} \frac{4\pi}{2l+1} \Ylm (\theta, \varphi)
  \\%
   \times \sum\nolimits_i q_i \Ylm^* (\theta_i, \varphi_i)
   \frac{r_{i,<}^l}{r_{i,>}^{l+1}}\, ,
\end{multline}%
and therefore, the potential inside the exclusion sphere arising from the
external screening ions is merely
\begin{equation}
  \tilde{\Phi}_<  (\vecr; \{\vecr_i, q_i \}) = \sum_{\lm} \frac{4 \pi}{2l+1}
  \Alm \, r^l\,  \Ylm (\theta,\varphi)\, ,
\end{equation}
where $\vecr = (r, \theta, \varphi)$ and the $\Alm$ are
given by \eqref{alm}.
The electrostatic contribution of the species $\s$ to the
total free energy now follows via the Debye charging process
\cite{deby&huck23,mef&levi93} as
\begin{equation}
  \label{free-nrg1}
  F^{\El}_\s(T,\{\rho_\s\},V)=
   \frac{N_\sigma}{D} \int_0^1 \sum\nolimits_i q_i d\lambda \,
   {\tilde \Phi_<}
  (\vecr_i; \{ \vecr_j, \lambda q_j \})\, ,
\end{equation}
and normalizing by $V\kb T$, we finally obtain
\begin{align}
  \label{free-nrg2}
  \barf^{\El}_\s(T,\{\rho_\s\}) & \equiv - F^{\El}_\s/V\kb T \nonumber \\
  & = \frac{\beta}{D} \rho_\s \sum_{l=0}^\infty \frac{4 \pi \, v_{2l}(x_\s)}
  {(2l+1) a_\s^{2l+1}} \, \sum_{m=-l}^{l}|\Qlms|^2\, ,
\end{align}
where the crucial expressions are
\begin{multline}%
  \label{omega}
   v_{2l}(x)= \int_0^1 d\lambda\, \lambda\, \left[1 -
    \frac{(2l+1) k_l(\lambda x)}{\lambda x \, k_{l+1}(\lambda x)} \right]
  \\
  = \frac{2l+1}{x^2} \left\{ \ln \left[
    \frac{g_{l+1}(x)}{g_{l+1}(0)} \right] -x +\frac{x^2}{2(2l+1)} \right\}
        \, ,
\end{multline}%
while the `multipole-squared amplitudes',
$\sum_m |\Qlms|^2$, are independent of the axes defining the polar
coordinates.


\subsection{Monomers}

Consider a monomer (or single $+$ or $-$ ion) with diameter
$a_\pm\pe a$ and charge $q_\pm$. The multipole expansion \eqref{qlm}
contains only the $l\pe 0$ term with $Q_{0,0}\pe  q_\pm /\sqrt{4 \pi}$.
Substituting into \eqref{free-nrg2} gives the  reduced free energy
of a monomer in a cloud of screening ions
\begin{equation}
  \barf^{\El}_\pm(T,\{\rho_\s\})=  \frac{q_\pm^2}{q_+ |q_-| T^*} \rho_\pm
  v_0(\kappa a)\, .
\end{equation}
If only monomers are present, summing the
contributions from the positive and negative ions leads to the
familiar DH free energy \cite{deby&huck23,mef&levi93}, namely,
\begin{equation}
  \label{fdh}
  {\bar f}^{\nDH}(T,\{\rho_\s\})=[\ln(1+\kappa a)-\kappa a+
  \mbox{$\frac{1}{2}(\kappa a)$}^2]/4 \pi a^3\, .
\end{equation}
This result, which depends
only on $x\pe \kappa a$ is, in fact, generally valid for any number
of charged species, provided all of them have the same size and
the system is overall electrically neutral: as well known, it
reproduces the exactly known answers at the leading low-density order.


\subsection{Dimers}

For our $z$:1 system, consider the dimer illustrated in Fig.~\ref{fig-dimer}
with a cation of charge $q_+ \pe z q_0$, separated from an anion of
charge $q_- \pe -q_0$ by a distance $a_1$. (In reality, $a_1 \ge a$
will be a fluctuating distance; but, as discussed in detail in
\textbf{I}, we may, in reasonable approximation, regard it as a
definite function of $T$: see also in Sec.~\ref{geom} below.) Let $a_2$
be the radius of the effective exclusion sphere (which, clearly,
should increase when $a_1$ increases).  Since the dimer is
asymmetric unless $z \pe 1$ we displace the center of the
exclusion sphere towards the positive cation, by a distance $p a_1$: see
Fig.~\ref{fig-dimer}. One should expect
the optimal value of $p$ to depend on $z$: by symmetry, one must
surely choose $p\pe \frac{1}{2}$ for $z\pe 1$, but when
$z \rightarrow \infty$
one should, likewise, have $p \rightarrow 0$. For the moment it
suffices to assume $0 \le p \le 1$: a concrete criterion for
choosing $p$ will emerge below.

In the configuration of Fig.~\ref{fig-dimer}, the leading multipole moments
are
\begin{equation}
  Q_{l,0} = \sqrt{\frac{2l+1}{4 \pi}} q_0 a_1^l [ z p^l + (-1)^{l+1}
    (1-p)^l ]\, ,
\end{equation}
and $\Qlm \pe  0$ if $m \neq 0$.
Substituting the above into \eqref{free-nrg2} yields the dimer
contribution
\begin{multline}%
  \label{f-dimer}
   \barf^{\El}_2(T;\{\rho_\s\}) = \frac{\rho_2}{z T^*}
   \times \\
   \sum_{l=0}^\infty \frac{a \, a_1^{2l}}{a_2^{2l+1}}
        [z p^l + (-1)^{l+1} (1-p)^l]^2 \, v_{2l} (x_2)\, .
\end{multline}%
The value of this sum and its rate of convergence clearly depends
on the value of $p$. Explicit numerical tests using $a_1/a \pe 1$,
$a_2/a \pe 3[1+\ln(3)/2]/4$ (the `angular average' value discussed in
Sec.~\ref{geom}) show that the series converges sufficiently rapidly that, to
the precision of interest, one need not consider terms beyond
$l\pe 2$ (see also \textbf{I}). Indeed, for $z\pe 2$ and $3$ and
$1 \leq x_2 \leq 5$, the
$l \! \geq \!3$ remainder for $p \pe \frac{1}{2}$
varies only from $0.8\%$ to $1.6 \%$ of the $l\pe2$ or dipolar term
and can thus be safely neglected within the
accuracy of this calculation. Evidently,
a reasonable criterion for the optimal value of $p$ would be that
which minimizes the full sum of terms from $l\pe 2$ to $\infty$. In
view of the rapid convergence, however, a very satisfactory option
is to choose the value of $p$ that minimizes the $l\pe 2$ or
quadrupolar term: this yields the simple result
\begin{equation}
  \label{choice}
  p={1 / (1+ {\sqrt {z}})} \, .
\end{equation}
This value,
in fact, eliminates the quadrupolar term entirely and satisfies
the two limiting cases, $z\pe 1$ and $z \rightarrow \infty$,
discussed above. Adopting this expression for $p$ and neglecting
the terms with $l \ge 3$ in \eqref{f-dimer}, we obtain the very
satisfactory approximation
 \begin{multline}%
  \label{f-dimer2}
   \barf^{\El}_2(T;\{\rho_\s\})= \frac{(z-1)^2}{z T^*}\, \rho_2
   \frac{a}{a_2}
    v_0(x_2) \, \, +
  \\
    \frac{1}{T^*} \rho_2 \frac{a a_1^2}{a_2^3}  v_2(x_2)\, ,
\end{multline}%
which we will employ below.  It is interesting to note
that the choice \eqref{choice} also makes the coefficient of the $l\pe 1$ (or
dipolar) term independent of $z$.  Furthermore, numerical
calculations show that for this choice of $p$, the $l \ge 3$
remainder divided by the $l\pe 1$ term
is reduced by a factor of about $1/z$ relative to the
symmetrical assignment $p\pe \frac{1}{2}$.


\subsection{Trimers}

\begin{figure}[t]
\scalebox{0.47}{\includegraphics{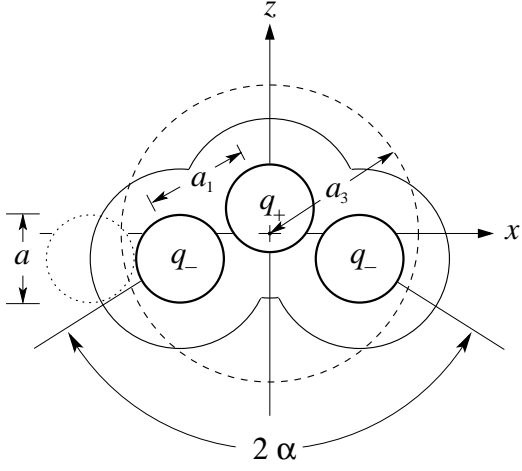}}
  \caption{\label{fig-trimer} A trimer bent at an angle $2\alpha$. 
    The dotted sphere indicates the closest approach by a screening ion.
    The exclusion zone is approximated by a sphere, shown dashed, of 
    radius $a_3$.}
\end{figure}

In considering the solvation of a trimer species, the first point
to note is that although the ground state is linear (in the form:
$-q_{0},+zq_{0},-q_{0}$) and so has a vanishing dipole moment, the
typical fluctuating configuration at finite temperatures must be
\textit{bent} and hence have a dipole moment of magnitude of order
$q_{0}a$.  Indeed, examination of snapshots of simulations for $z
\ge 2$ in the critical vicinity (see, e.g., \cite{pana&mef02}) fully
confirms this conclusion.  Accordingly, consider, as illustrated
in Fig.~\ref{fig-trimer}, a trimer which is (say, ``instantaneously'') bent 
at an
angle $2\alpha$.  To simplify the analysis, we will suppose that
it is adequate to fix the radial distances $r_1$ and $r_2$ for
both satellite anions at the spacing $a_1$.  (As discussed in
$\bf{I}$, and also below, we expect the fluctuations in $a_1$ to
be relatively small.)

For the effective exclusion sphere, now of radius, say, $a_3$ (see
Fig.~\ref{fig-trimer}), 
the issue of the placement of its center again arises. By
symmetry (having imposed $r_{1}\pe r_{2}\pe a_{1}$) the center should
lie on the bisector of the angle $2\alpha$ which, in Fig.~\ref{fig-trimer}, 
has
been identified as the $z$-axis.  Then, in analogy to the dimer,
we center the exclusion sphere at a distance
$pa_{1}\textrm{cos}\alpha$ displaced from the center of the
central cation (or charge $q_+$) towards the two anions of charge
$q_{-}$ whose axial location  lies at a distance $a_{1} \cos \alpha$
as projected onto the bisecting axis. With
this placement of the center, we find the multipole-squared amplitudes
\begin{align}
  |Q_{0,0}|^2 & = (1/4\pi) (z-2)^2 q_0^2 \, , \\
  \sum\nolimits_m |Q_{1,m}|^2 & = 
    (3/4\pi) [zp+2(1-p)]^2 (\cos \alpha)^2 q_0^2 a_1^2 \, , \\
  \sum\nolimits_m |Q_{2,m}|^2 & = 
  (5/4\pi) \biglb \{ 3 \sin^4 \alpha  + \biglb(\sin^2 \alpha
     \nonumber \\%
     & \phantom{=} +   [zp^2 - 2(1-p)^2] \cos^2 \alpha \bigrb)^2 \bigrb\} 
     q_0^2   a_1^4 \, .
\end{align}


In an ideal calculation of the solvation free energy of trimers,
every trimer bent at a specific angle would be treated as a
separate species in its own right.  However, to make our
calculations tractable we substitute these expressions for the
multipole-squared moments into the basic result \eqref{free-nrg2} and
replace the factors that depend on $\alpha$ by thermal averages to
obtain
\begin{widetext}
\begin{multline}
  \label{f-trimer}
  \barf^{\El}_3(T;\{\rho_\s\}) = \frac{\rho_3}{z T^*}
  \left\{(z-2)^2 \frac{a}{a_3} \, v_0(x_3) \right. 
    +[z p +2(1-p)]^2 \, \left\langle {\cos^2 \alpha} \right\rangle \, 
         \frac{a a_1^2}{a_3^3} \, v_2(x_3) \\
  \left. + \left\langle 3 \, {\sin^4 \alpha} + \{ [z p^2 -2(1-p)^2] 
         \cos^2 \alpha + \sin^2 \alpha \}^2 \right \rangle \, 
         \frac{a a_1^4}{a_3^5}\, v_4(x_3) + \ldots 
   \right\} \, .
\end{multline}
\end{widetext}

Again, an ideal calculation would recognize that the increased
solvation free energies resulting from larger dipole moments,
should enhance the thermal weight of more highly bent trimers.
However, we will forgo such a refinement (which would require a
cumbersome self-consistent formulation) and merely weight the bent
trimer configurations via the Boltzmann factors computed with the
``bare'' cluster energies.  Accordingly, we consider the 
thermal average $\langle \cal O \rangle$ of an angular function
$\cal O$ at temperature $T$ to be defined by
\begin{multline}%
  \label{scriptO}
  \langle
  \mathcal{O} \rangle \equiv \int_{\pi / 6}^{\pi / 2} d\alpha \, \sin 2
  \alpha\, \mathcal{O} (\alpha) \, e^{E(\alpha)/\Te} \biggl /
  \\%
  \int_{\pi / 6}^{\pi / 2} d\alpha\,  \sin 2\alpha
  \, e^{E(\alpha)/\Te}\, ,
\end{multline}
where $E(\alpha) \pe  -a/2z a_1 \sin {\alpha}$
is the reduced repulsion energy between the two
satellite anions. Note that in setting the lower limits of
integration at $\alpha\pe \pi / 6$ (due to hard-core repulsions),
we have neglected a domain of
closest approach, and, hence, highest repulsive energy, that is
accessible when $a_1 > a$. In fact, in the following calculations,
we will need only the two averages
$\langle \sin^2 \!\alpha \rangle $ and $\langle \sin^4 \! \alpha \rangle $
which follow from the expressions
\begin{equation}
  \label{}
  \langle \sin^{2n} \! \alpha \rangle = \frac{1}{(2z\Te)^{2n}}
  \frac{s_{2n+3}(1/z\Te) - s_{2n+3}(1/2z\Te)}
     {s_{3}(1/z\Te) - s_{3}(1/2z\Te)}\, ,
\end{equation}
where 
\begin{equation}
  \label{}
  s_n (x) = \frac{(-1)^{n-1}}{(n-1)!} \left[ \textrm{Ei} (-x) 
  + \frac{e^{-x}}{x} \sum_{k=0}^{n-2} (-1)^k \frac{k!}{x^k} \right]\, .
\end{equation}

As regards the choice of $p$, a first guess is to
choose the value $p_{\min}$ which minimizes the quadrupolar term. 
However, this leads to unphysical features such as
$p_{\min} < 0$ and even to $p_{\min} \rightarrow - \infty$
when $\Te \rightarrow 0$ (when, in fact, trimers become straighter and 
straighter).  
The alternative adopted here is to accept the value of $p$ in the interval 
$0 \le p \le 1$ that minimizes the quadrupolar term. One may verify 
that this value is $p\pe 0$ for  
$0.003 < \Te <  0.06$, yielding a quadrupolar term 
that agrees  with the exact minimum to within $3\%$.



In summary, although, as indicated, various refinements of our
approach may readily be contemplated, we believe that the
formulation reasonably captures the essential physics underlying
the solvation of fluctuating trimeric ion clusters.


\subsection{Tetramers}

For $z \ge 3$, one must allow for the formation of tetramers and
include their solvation free energy.  A tetramer in its ground
state is planar with $\varphi\pe 0$, satellite radii
$r_i\pe |{\bf{r}}_i|\pe a$, 
$(i\pe 1,2,3)$ and angular separations,
$\theta_{12}\pe \theta_{13}\pe 2\pi/3$ (see Fig.~\ref{fig-tetramer}). 
As for the trimers,
thermal fluctuations about the ground state configuration give
rise to significant dipole moments that are absent at $T\pe 0$. To
tackle this issue we estimate the solvation free energy of a
tetramer by considering the harmonic normal modes of \textit{angular}
oscillation about the ground state configuration.  (Note that
these modes already enter into the calculation of the
corresponding association constant:  see Appendix \ref{appa}.)  For each
mode, a thermal average of the contributions of the individual
multipole moments is computed; the sum of these mean-square
terms then provides a value for the overall multipole free energy.
Of course, this approximation neglects the nonlinear interactions
between the modes but, because of the relatively low value of the
critical temperature, this should not be numerically significant.

Following our treatment of trimers, we will fix the three
satellite radii at $a_1$; the exclusion zone for the tetramer will
be approximated by a sphere of fixed radius $a_4$ which we choose
to center on the positive core ion (of charge $+zq_0$). Ideally,
the origin of the effective exclusion sphere should again be
placed so that, say, the total contribution of the quadrupolar
free-energy term is minimized.  For the present calculations,
however, only the case $z\pe 3$ will be utilized:  then the tetramers
are neutral so that both monopole and dipole moments are
independent of the origin about which they are defined.  The
variations of the quadrupole moments due to small displacements of
the origin and, likewise, variations in the exclusion diameter
$a_4$ that might reasonably be associated with thermally induced
shape changes, may be neglected at the level of precision
appropriate in light of the other approximations of the theory.
(Note that changes in the definition of $a_4$ are studied
quantitatively in Sec.~\ref{results}, below.)

A planar tetramer has three angular normal modes: two `in-plane'
modes and one `out-of-plane' mode.  The first two modes correspond
to $\varphi\pe 0$ in Fig.~\ref{fig-tetramer} 
[and $\varphi^*\pe 0$ in \eqref{app4} of Appendix \ref{appa}]
since the three satellite ions remain in the $(x,y)$ plane and ion
1 may be considered as fixed on the $x$-axis (at $x_1\pe r_1\pe a_1$). The
first mode ($a$) is a `flapping' mode in which the ions 2 and 3 (see
Fig.~\ref{fig-tetramer}) oscillate in phase, towards and away from the axis 
formed by ion 1 and the central, positive ion; in other words one has
$\theta_{12} \pe \theta_{13}$ [following from $Y \pe 0$ in \eqref{app4}].  
The
second mode ($b$) is a `pendulum mode' in which the angle between
the ions 2 and 3 remains fixed, equal to its equilibrium value so
that $\theta_{12} + \theta_{13} \pe  4\pi/3$ {corresponding to $X\pe 0$
in \eqref{app4}]. Lastly, in the `out-of-plane' mode ($c$), two satellites 
are fixed whereas the third one swings around the plane of equilibrium
(corresponding to the mode where $X\pe Y\pe 0$ while $\varphi$ is varying) 
\footnote{For symmetry reasons, one might prefer to consider, in place of
the eigenmode ($c$), the more symmetric mode ($c'$) in  which the three 
satellites swing together in and out of the equilibrium plane while
maintaining an equilateral triangle geometry. For small deviations
around the ground state, these two modes coincide; furthermore the 
differences induced in the resulting critical temperatures and densities
(via the nonlinearities) are less then $0.3\%$.}.


For each mode we need a configuration-space weighting factor:
these all derive from the expression \eqref{assoc-const} for the association
constant.  For the tetramer, using the coordinates in
Fig.~\ref{fig-tetramer}, this is
\begin{multline}%
  d\textbf{r}_1 d\textbf{r}_2 d\textbf{r}_3 = 8\pi^2 {r_1}^2
  dr_1 dr_2 dr_3 
  \\
  \times
  ({r_2}^2\sin\theta_{12} \, d\theta_{12})({r_3}^2
  \sin\theta_{13} \, d\theta_{13})d\varphi \, ,
\end{multline}%
where a prefactor
$4\pi{r_1}^2$ comes from the full angular integral over the
orientation of the $x$ axis, while a factor $2\pi$ arises from the
axial integral (rotating the $y$ axis about the $x$ axis so that
satellite $2$ is in the $(x,y)$ plane). Insofar
as we consider the angular modes at fixed $r_i\pe a_1$, the only
relevant factor for the angular averages over the mode coordinates
is $\sin\theta_{12} \sin\theta_{13} \, d\theta_{12} \, d\theta_{23}
d\varphi$.  (Note that essentially identical considerations enter
in writing \eqref{scriptO} where $\theta_{12}\pe 2\alpha$.)

Now the monopole moment of the (general) tetramer is always
$Q_{0,0} \pe  (z-3)q_0 / \sqrt{4 \pi}$; but the higher moments clearly
depend on the mode configurations as we proceed to specify.

(a) \textit{\textbf{In-plane flapping mode}}.  Let
$\theta_{12}\pe \theta_{13}\pe \theta$ be the angle describing this
normal mode (but recall that $\theta \pe  2\pi / 3$ specifies the
ground state).  The dipole and quadrupole amplitudes generated by
excitation of the mode are then
\begin{align}
 \label{qlma}
   \sum\nolimits_m |Q_{1,m}|^2  & = \frac{3}{4\pi}
   [1+2 \cos \theta ]^2 q_0^2 a_1^2  \, ,\\
   \sum\nolimits_m |Q_{2,m}|^2  & = \frac{15}{4\pi}
   [\sin^4 \theta + 3 \cos^4 \theta ] q_0^2 a_1^4 \, .
\end{align}
The reduced  repulsive Coulombic energy between the three
satellite ions is given by
\begin{equation}
  {E_a(\theta)} = -\frac{a}{z a_1}
        \left[\frac{1}{\sin (\theta/2)} + \frac{1}{{2\sin\theta}}
        \right] \, .
\end{equation}
Thus the thermal average square moments may be calculated from
\begin{equation}
{\langle |\Qlm^2| \rangle }_a = \mathcal{N}_a(T^*) \int_{\pi /
3}^{5\pi / 6} d\theta \sin^2 \theta \, |\Qlm (\theta)|^2
\, e^{E_{a}(\theta)/T^*}\, ,
\end{equation}
where $1/\mathcal{N}_a(T^*)$ is the
obvious normalizing integral.  The limits specified on $\theta$
correspond to the hard-core restrictions in the
case $a_1 \pe  a$ and should be relaxed
appropriately if $a_1 \! > \!a$ (although, since they correspond to the
maximal interionic repulsions, the differences will be small).

(b) \textit{\textbf{In-plane pendulum mode}}: Now let us put
$\theta_{12}\pe (2\pi / 3)-\theta'$ and $\theta_{13}\pe (2\pi /3) +
\theta'$, so that $\theta'$ describes the angular
amplitude of the mode.  The dipole and quadrupole amplitudes are then
\begin{align}
   \sum\nolimits_m |Q_{1,m}|^2  &
   = \frac{3}{2\pi} [1 - \cos \theta'] q_0^2 a_1^2   \, ,\\
   \sum\nolimits_m |Q_{2,m}|^2  &
   = \frac{15}{16\pi} [5-2\cos(2 \theta')] q_0^2 a_1^4  \, .
\end{align}
The thermal average is now computed via  the normalized integration
\begin{multline}%
  \langle |\Qlm|^2 \rangle_b = \mathcal{N}_b (T^*) \int_{-\pi/3}^{\pi/3}
  d\theta' [1+2 \cos (2 \theta')] \,
        \times  \\
  |\Qlm(\theta')|^2
  \, e^{E_b(\theta')/T^*}\, ,
\end{multline}%
where the reduced energy can be written as
\begin{equation}
  E_b(\theta') = - \frac{a}{z a_1} \left[
  \frac{1}{\sqrt{3}} + \frac{2\sqrt{3}\cos (\theta'/2)}
  {1+2\cos\theta'}\right] .
\end{equation}

(c) \textit{\textbf{Out-of-plane mode.}}
Finally, in the out-of-plane mode as described previously (in which
$\theta_{12} \pe  \theta_{13} \pe  2\pi/3$ and $\varphi$ varies), the
dipolar and quadrupolar amplitudes are
\begin{align}
   \sum\nolimits_m |Q_{1,m}|^2  & 
   = \frac{9}{8\pi} [ 1 - \cos \varphi ] q_0^2 a_1^2  \, ,\\ 
   \sum\nolimits_m |Q_{2,m}|^2  & 
   = \frac{45}{64\pi} [3-2\cos \varphi + 3 \cos^2 \varphi] q_0^2 a_1^4\, ,
\end{align}
while the reduced repulsive energy is simply 
\begin{equation}
  \label{}
  E_c (\varphi) = - \frac{a}{\sqrt{3} z a_1} \left[ 2
        + \frac{1}{\cos (\varphi/2)} \right] \, .
\end{equation}
For this mode, the thermal average is performed according to
\begin{equation}
  \label{}
  {\langle |\Qlm|^2 \rangle}_c =   \mathcal{N}_c (\Te) \int_0^{\varphi_m}
  d \varphi \, |\Qlm(\varphi)|^2  e^{E_c (\varphi)} \, ,
\end{equation}
where the condition 
$\varphi \leq \varphi_m \equiv \pi \! - \! 2\arcsin (1/\sqrt{3})$ 
expresses the hard-core condition and $1/\mathcal{N}_c (\Te)$ is the 
corresponding normalizing integral.

At this point, the overall solvation free energy of a tetramer, 
$\barf^{\El}_4 (T; \{\rho_{\s}\})$, may be calculated by summing the 
solvation free energies computed for each mode. This completes the
basic general analysis.



\section{Criticality and Coexistence under Charge Asymmetry}
\label{para}

\subsection{Pure Debye-H\"uckel Theory}
\label{pure-DH}

The original Debye-H\"uckel theory \cite{deby&huck23} amounts to writing the
overall free energy density as 
\begin{equation}
  \label{f-dh}
  \barf(T,\rho_+,\rho_-)=\barf^{\nDH}(T,\rho_+,\rho_-)+\barf^{\Id}(T
  ,\rho_+)+\barf^{\Id}(T,\rho_-)\, , 
\end{equation}
 where $\rho_+$ and $\rho_-$ are
the densities of cations (with charge $q_+ \pe zq_0$) and anions
(with charge $q_- \pe -q_0$), respectively, and $\barf^{\nDH}$ was
obtained in \eqref{fdh}. From the electroneutrality
condition \eqref{electroneutrality} one has 
\begin{equation}
  \rho_+= \rho /(1+z), \quad \rho_- = z  \rho/(1+z) \, , 
\end{equation}
while using the
expression \eqref{kappa} for Debye length, with $x\!\equiv\! \kappa a$, 
the normalized density~\eqref{rho*} becomes
\begin{equation}
  \rho^*=x^2 T^*/4\pi\, . 
\end{equation}
The contribution to the overall chemical potential
from the DH free energy is then 
\begin{equation}
  \label{mu-dh} 
  \barmu^{\nDH}=  -{x / 2 T^*(1+x)}\, . 
\end{equation}
Taking $\mathcal{C}_+ \pe \mathcal{C}_- \pe \Lambda_1^3$
in~\eqref{id-nrg} where $\Lambda_1(T)$ is the de Broglie
wavelength for free ions, the ideal gas contribution is merely
 \begin{multline}%
  \label{muid}
  \barmu^{\Id} =\ln (x^2 T^* ) + \frac{1}{1+z}\ln \left(
        \frac{1}{1+z}\right)
      \\%
     + \frac{z}{1+z} \ln\left(\frac{z}{1+z}\right)+ \ln
  \left( \frac{\Lambda_1^3}{4 \pi a^3}\right)\, .
\end{multline}%
The overall chemical potential is
$\barmu \pe \barmu^{\nDH}+\barmu^{\Id}$, while the reduced pressure
follows from \eqref{pressure} as
\begin{equation}
  p^{\dagger} \equiv 4\pi a^3 \barp = x^2
  T^* + \ln(1+x)-x + \mbox{$\frac{1}{2}$}x^2/(1+x)\, ,
\end{equation}
which is the
same as (4.6) in {\bI} and quite independent of $z$.

Since the expression for the pressure does not depend on $z$ and
the overall chemical potential is also $z$-independent except for
the constant terms in the ideal gas form~\eqref{muid}, the
conditions for criticality and phase coexistence are identical to
those derived in {\bI} for the 1:1 model. The phase coexistence
curves are likewise identical: see {\bI} Fig.~1(a).  In
summary, the pure Debye-H\"uckel theory predicts that the critical
parameters are \textit{independent} of $z$ and given by
 \begin{multline}%
    T_c^*=1/16,\quad \rho_c^*=1/64\pi, \quad x_c=1, \quad
  \\%
    Z_c \equiv p_c/\rho_c \kb  T_c=16 \, \ln 2 - 11\, ,
\end{multline}%
while the numerical values are presented in Table \ref{table}.


\subsection{DHBjCIHC Theory}

Extending the DHBjDIHC
pairing-plus-solvation approach for 1:1 electrolytes 
to $z$:1 electrolytes, we now include dimers, trimers, and all
further primary clusters up to $(z+1)$-mers, and add their
free energies to the overall electrostatic free energy to obtain 
\begin{equation} 
  \barf(T;\{\rho_\s\}) = \sum_\nu \left[
    \barf^{\Id}(\rho_\nu)+\barf^{\El}_\nu(T;\{\rho_\s\}) \right] \, , 
\end{equation}
where
$\nu\pe +, -, 2, 3 \ldots$ for positive ions, negative ions, dimers,
trimers, $\ldots$, respectively.  To determine the degree of
association of the free ions into dimers, trimers, $\ldots$, we
need the association constants $K_{m,z}(\Te)$ as
computed in Sec.~\ref{assoc}. Then, under chemical equilibrium the cluster
densities $\rho_m$ for $2\le m \le z+1$, satisfy the laws of mass
action in the form
\begin{equation}
  \label{mass-action11}
  \rho_m = K_{m-1,z} \, \rho_+ \rho_-^{m-1} \exp[\mu^{\El}_+ + (m-1) 
  \mu^{\El}_-  - \mu^{\El}_m ] \, ,
\end{equation}
where the excess chemical potentials are given from {\bI} by 
\begin{equation}
  \mu^{\El}_\nu (T;\{\rho_\s\}) \equiv - \left. \partial {\bar
  f}^{\El}/\partial \rho_\nu  \right|_{T,\rho_{\s'}}  \, .
\end{equation}                    
We dub this extended
treatment DHBjCIHC theory for ``\textbf{D}ebye-\textbf{H}\"uckel theory
supplemented by \textbf{Bj}errum association into \textbf{C}lusters that
are solvated by the \textbf{I}onic fluid, and \textbf{H}ard \textbf{C}ores.''

In order to obtain a canonical equilibrium state of the 
$(z\!+\!2)$-component 
fluid, one needs, in addition to electroneutrality and the $z$ 
mass action conditions, one extra
parameter, such as the overall density $\rho$ or, more conveniently, 
the reduced Debye variable $x \pe \kappa a$. 
Moreover, phase coexistence entails 
the conditions \eqref{equilibrium} and \eqref{eq-galvani}, 
whereby one can show that the equality of all the different 
electrochemical potentials between coexisting phases
can be replaced 
by the  equality of the chemical potential of the neutral species 
\textit{alone}
(dimer, trimer, or tetramer, respectively, for $z\pe 1$, $2$, or $3$).
(For the charged species, the 
\textit{electro}chemical potential must match between two 
coexisting phases as mentioned in the Introduction and discussed in
Sec.~\ref{galvani} below in connection with the Galvani potential.) 
One effective computational strategy is thus to plot
\textit{parametrically} $\bigl(\bar{p}(x,T), \bar{\mu}_n (x,T)\bigr)$
(where $n$ denotes the
neutral species), and to seek  for two different values of $x$ giving
the same point: see Fig.~\ref{fig-pmu}.
Especial care is needed in determining the coexistence curve below
criticality for $z \pe  3$.

\begin{figure}[t]
\scalebox{0.43}{\includegraphics{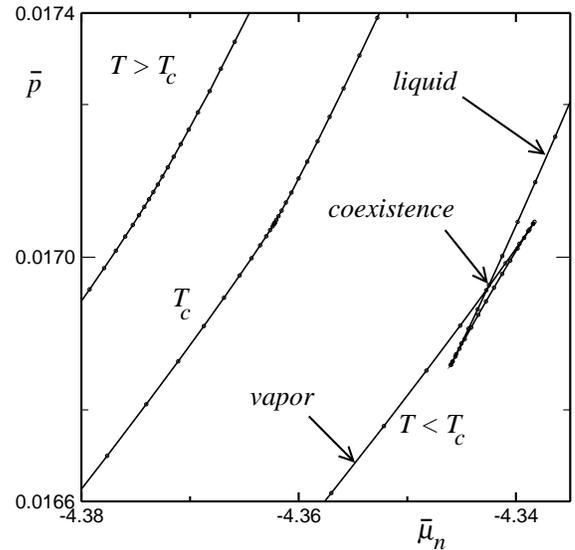}}
  \caption{\label{fig-pmu}
    Examples of the variation of the pressure with
    the chemical
    potential of the neutral species $\mu_n$ (shifted by an arbitrary
    constant), calculated for a $3$:$1$ electrolyte treated
    within the DHBjCIHC theory with refined standard parameters.
    Two-phase coexistence below $T_c$ can be realized
    when the curve intersects
    itself (while, as usual, the states below the intersection
    are not stable). The plots are constructed
    parametrically as functions of $x \pe \kappa a$ using
    increments of $0.03$ around $x_c \pe 1.570$ for reduced temperatures
    $T^* \pe 0.04250$,
    $0.043345$ and $0.04380$.}
\end{figure}

However, calculations in the single-phase region (above
$T_c$) are relatively straightforward;
consequently, for the purpose of calculating $T_c$ and $\rho_c$ another useful
approach is to generate supercritical loci which must
intersect at the critical point.
We choose the maxima of the $k$-susceptibilities (see
Sec.~\ref{chik}), which, indeed, lead to fast and accurate
determinations of $T_c$ and $\rho_c$.


\subsection{Geometric parameters for the ion clusters}
\label{geom}

To proceed further in the quantitative evaluation of the electrostatic
contributions to the free energy as derived analytically in the
previous sections, we must address the values and thermal
variations of the satellite separation radii, $a_1$ (which should both
depend
on the cluster species $m$ and the valence $z$) and of the
effective exclusion sphere diameters $a_m$ for $m\geq 2$. Both these
issues were discussed for the basic 1:1 model (the RPM) in {\bI} (see
Secs. 6.3 and 7.1) and so will be treated fairly briefly here.

If we write $a_1\pe a[1+s_{1,m,z} (T)]$, it is, first, clear that
$s_{1,m,z}$ vanishes when $T\rightarrow 0$ for all $m$ and $z$, so that
hard-core contact is, in fact, rather rapidly approached when $T$ falls.
Indeed, for $\Te \leq 0.055$ [$\simeq \! \Tce (z\pe 1)$], the analysis
of {\bI} indicates that $s_{1,1,1}$ decreases almost linearly with $\Te$
from $s_{1,1,1} \simeq 0.08$. When $z\!>\!1$,
because of the tighter binding induced by the
larger central charges, which is only partly offset by
repulsions from the remaining $m-1$ satellite ions, one must also have
$s_{1,m,z} (\Te) < s_{1,m,z'} (\Te)$ when $z' >z$.

Then, one should also observe
(see Figs.~\ref{fig-Tc2z}, \ref{fig-rc2z} and Table \ref{table})
that for larger values 
of $z$ ($\geq 2$), the critical temperatures fall, so that the 
relevant values of $s_{1,m,z} (\Te)$ will again be smaller than for the
1:1 model. Nevertheless, one must notice from \eqref{free-nrg2}, 
\eqref{f-dimer2}, \eqref{f-trimer} and \eqref{qlma}, etc., 
that the dipolar and quadrupolar contributions (when the 
latter do not vanish by choice of the parameter $p$) are proportional
to $a_1^2$ and $a_1^4$, respectively. However, in compensation, these powers
are always accompanied by the \textit{inverse} powers $a_m^{-1}$ and
$a_m^{-3}$, respectively, of the exclusion diameters $a_m$ ($m \geq 2$),
which are proportional to the corresponding
values of $a_1 (T)$ and so act to reduce the overall sensitivity.

In {\bI}, the choice of the radius $a_2$ for the effective exclusion sphere
that approximates the true bispherical exclusion zone of a dipolar dimer
(see Fig.~\ref{fig-dimer}) was discussed by considering various bounds and
their mean values. It was decided to accept, as most appropriate, the
`\textit{angular average}' value, defined as the radius averaged over solid
angle of the true exclusion zone as measured from a \textit{symmetrically
located origin} of a cluster in its ground state. For dimers, trimers, and
tetramers in their ground states, these angular averages are, respectively,
\begin{align}
  \label{angav}
        \frac{a_2^a}{a} & = \frac{3}{4} + \frac{3}{8} \ln 3 , \quad
             \frac{a^a_3}{a} = \frac{5}{4}, \quad
        \frac{a^a_4}{a} = \frac{11}{8},
   \nonumber \\
 & \simeq 1.16198,  \quad
  \phantom{a^a_3/a \hspace{1mm} } = 1.25 , \quad
 \phantom{a^a_4/a \hspace{-6mm}} =  1.375 \, .
\end{align}

It transpires in the calculations leading to the critical parameters,
that the predicted values for all three cases, $z \pe 1$, $2$, and $3$,
are dominated by the properties of the \textit{primary neutral clusters},
namely, the neutral dimers, trimers, and tetramers, which prove to be
by far the most abundant species. In turn, for fixed $z$, these are found to
be the most sensitive to the geometrical parameters. Accordingly, we have
examined (as, in fact, did Levin and Fisher) various other more-or-less
plausible criteria. One simple, but clearly rather arbitrary possibility, is
to choose $a_{\s}$ so that the approximating exclusion sphere has a volume
matching that of the true exclusion zone. We identify these parameters as
`steric': they take the values
\begin{align}
  \label{steric}
  \frac{a_2^s}{a} & =   \frac{3}{2^{4/3}}, \quad   \frac{a_3^s}{a}  = \frac{19^{1/3}}{2}, \quad
  \frac{a_4^s}{a} = \frac{7^{2/3}}{2^{4/3}} \, , \nonumber \\
 &  \simeq 1.19055,
 \hspace{4mm} \simeq 1.33420, \hspace{6mm}
 \simeq 1.45220 \, .
\end{align}

Another choice, since the interactions that are being truncated by the
exclusion zones are Coulombic, is the harmonic diameters defined
as the \textit{inverse} of the angular average (again taken from
the clusters geometric center of symmetry) of the \textit{inverse
radial distance} to the surface of the exclusion zone. This leads to the
values
\begin{widetext}
\begin{align}
  \label{harmonic}
    \frac{a_2^h}{a} & =  \frac{6}{(2+3\ln 3)}, \quad
    \frac{a_3^h}{a}  \pe \frac{2}{(1+\ln 2)} , \quad
    \frac{a_4^h}{a}  \pe \frac{4}{(1+3\ln 2)}, \nonumber \\
 &  \simeq 1.13297,
 \hspace{1.4cm} \simeq 1.18123, \hspace{1.1cm} \simeq 1.29894 \, .
\end{align}
\end{widetext}
Compared to the angular averages \eqref{angav}, these
are some $2.5-5.6\%$ lower  which leads to
increased solvation. While, in accord with {\bI}, we judge that
the angular averages are to be preferred, the predictions of the
steric and harmonic parameters will be discussed below.

In as far as the satellite separation $a_1 (T)$ exhibits a $T$-dependence,
this will be inherited by the $a_\sigma (T)$.
However,
in the case of charged asymmetric dimers, as needed for $z \geq 2$, the
offset of the center of the effective exclusion sphere from the
clusters' geometric center [as embodied in \eqref{choice}] naturally raises
the question: Why not calculate the $a_{\s}$'s from the offset center ? 
Likewise, at finite temperature, the crucial bending fluctuations of the 
trimers and tetramers obviously suggest further modifications in the 
calculation of the $a_\sigma$'s. The temptation to explore these 
refinements however, may be resisted, first, because the effects 
are likely to be small, and, just as important, because the 
resulting changes in critical parameter estimates will be less
significant than result from other approximations already accepted.



\section{Quantitative predictions}
\label{results}

At this point, it is imperative to re-emphasize that the 
primary aim of the present study is to elucidate the basic 
physical mechanisms underlying the systematic trends in the 
various critical parameters that are induced as $z$ increases, and,
at a semiquantitative level, to understand the magnitudes of the 
changes. Recall that the true values of $\Tce (z)$, etc., are already 
known to satisfactory accuracy from the recent simulations 
\cite{young&mef04,pana&mef02}. 
Consequently, a \textit{uniform} theoretical treatment of the 
1:1, 2:1, and 3:1 models is of greater importance than are concerns for 
various specific subtleties that we know, \textit{a priori}, cannot 
yield truly reliable and accurate critical-point data owing to our failure 
(not to say inability) to treat adequately the essential critical 
fluctuations:
see, e.g., \cite{mef&lee96}. The fluctuations, of course, serve to
realize the universality class of the critical behavior
\cite{mef1,stell1,luij&mef02,young&mef&luij03,young&mef04,mef&lee96} 
while, at the same time, 
depressing the critical temperature and (for these primitive 
electrolyte models) increasing the critical density relative to the 
predictions of even ``the best,'' classical mean-field, or self-consistent 
treatments. 

With these points in mind, the principal explicit numerical calculations 
of the electrostatic free energy terms that we have undertaken have 
utilized the simple ($T\pe 0$) angular averages diameters $\as$, listed 
in \eqref{angav} and, furthermore, have accepted the ``in-contact'' or 
$T \rightarrow 0$ limit, $a_1 \pe  a$, for the satellite ion separations in 
all clusters. It should be stressed, however, that the calculations
of the cluster association constants, $K_{m,z} (T)$ in Sec.~\ref{assoc} are 
\textit{not} so constrained: rather, each satellite ion is allowed to 
explore the full phase space restricted only, at large separations, 
by the Bjerrum-type optimal cutoffs, $R_{m,z}$. 

\begin{table}
\caption{\label{table} Predicted critical parameters, 
  $\Tce \pe D a \kb T_c / z q_0^2$, $\roce \pe \rho_c a^3$, 
  $x_c \pe \kappa_c a$, 
  $Z_c \pe p_c / \rho_c \kb T_c$ and the mole fraction of free ions,  
  $y_{\pm \, c} \pe (N_+ + N_-)/N|_c$, 
  for $z$:1 hard sphere electrolytes, as predicted by the DHBjCIHC theory 
  with `standard' parameters (refined for $z \pe3$): see text.
  Monte Carlo results 
  \cite{young&mef04,pana&mef02} are displayed in parentheses.}
\begin{ruledtabular}
\begin{tabular}{llllll}
 $z$ & $10^2 \, \Tce$ & $10^2 \, \roce$ & $x_c$ & $Z_c$ & $y_{\pm \, c}$ \\
 DH & $6.250$ & $0.4974$ & $1$ & $0.9063$ & $1$  \\
$ 1$ &$ 5.567$ ($4.93_3$) & $2.614$ ($7.50$) & 1.038
&0.2451 & 0.1828 \\
$ 2$ &$ 4.907$ ($4.70$) & $6.261$  ($9.3$) & 1.366
 & 0.1708 & 0.1164 \\
$ 3$ &$ 4.334$ ($4.10$) & $11.90$ ($12.5$) & 1.570
& 0.1433  & 0.0838 
\end{tabular}
\end{ruledtabular}
\end{table}


\subsection{1:1 or Restricted Primitive Model Electrolyte}

Here, we merely refine the results of Fisher and Levin from {\bI}. 
Within the DHBjCI theory using the angular average for 
$a_2$ (but \textit{without} the contribution 
$\barf^{\HC}$), one finds
\begin{equation}
  \label{su}
  \Tce \pe  0.05740 \quad \mbox{and} \quad  \roce \pe  0.02779 \, . 
\end{equation}
We may supplement the results of 
{\bI} by recording that the use of the larger \textit{steric} 
parameter $a_2^s$ [see \eqref{steric}] modifies
the predictions for $\Tce$ and $\roce$ by factors $0.9815$ and $1.0086$, 
respectively,  
whereas, the smaller \textit{harmonic} exclusion diameter $a_2^h$,
yields factors $1.0195$ and $0.9989$. 
Hence, a larger cluster size leads naturally  to a decrease in 
$\Tce$, since fewer attractions are realized, and to an increase in $\roc$.

The next step is to include the hard-core term 
$\barf^{\HC}$. Keeping the angular average radius $a_2^a$ and taking the  
bcc hard-core value $B_\s / a_\s^3 \pe  4/3\sqrt{3}$, a choice of parameters
that we will refer to as \textit{`standard'}, we find 
the critical parameters displayed in Table \ref{table}. 
As expected, the introduction of hard-cores reduces \textit{both} the 
critical temperature (by around $3\%$) and the critical 
density (by $6\%$). These effects are stronger if the 
low-density limiting value 
$B_\s / a_\s^3 \pe  2\pi /3$ is used since $\Tce$ then drops to 
$0.05293$, i.e. by $5\%$, 
whereas $\roce$ becomes $0.02469$, falling by $6\%$. Finally, using the
angular average $a_2^a$ but with the   
choice $B_\s / a_\s^3\pe  1.300$, which lies between the 
low-density and bcc values, we obtain the `optimal-fit' estimates
\begin{equation}
  \label{sd}
  \Tce (z \pe 1) \pe 0.05455 \quad \mbox{and} \quad 
  \roce(z \pe 1) \pe 0.02542 \, .
\end{equation}

The coexistence curves predicted by the DHBjCI theory (with the angular 
average for $a_2$) and by the DHBjCIHC theory with standard parameters  
are plotted in Fig.~\ref{fig-diagphas}. The introduction 
of $\barf^{\HC}$ significantly lowers the liquid sides of the coexistence 
curves. One may notice 
that the Monte Carlo data could well be better fitted by some choice 
of $B_\s$ between $0$ and the bcc value. 

Note that in addition to the `standard' values of $\Tce$ and $\roce$, 
listed in Table 
\ref{table}, the last column, labeled $y_{\pm \, c}$, reports the 
critical value of the \textit{mole fraction of unassociated ions}, 
namely, 
\begin{equation}
  \label{}
  y_\pm \pe  y_+ + y_-, \quad \mbox{with} \quad y_\sigma \pe n_\s N_\sigma 
  / N \, ,
\end{equation}
where $y_\s$ is the
mole fraction of species $\s$ while $n_\s$ is its ionic weight (i.e.,
$n_\s \pe 1$ for free ions but $n_\s \pe m$ for a cluster of one positive
charge and $(m-1)$ negative charges). For the 1:1 model, we have
$y_{+ \, c} \pe  y_{- \, c} \pe  0.0914$ while the critical mole fraction
of the \textit{associated} ion pairs is
$y_{2 \, c} \pe  2 \rho_{2 \, c} / \rho \pe  0.8172$: see Table
\ref{table-ys}.
The fact that (within
the DHBjCIHC theory) almost $80\%$ of the ions are associated into dipolar
ion pairs near criticality makes it less surprising that a model of neutral
but charged hard dumb-bells might have a comparable coexistence curve, as
some simulations suggest \cite{shel&pate95,daub&pate03}.

\begin{figure}[t]
\scalebox{0.42}{\includegraphics{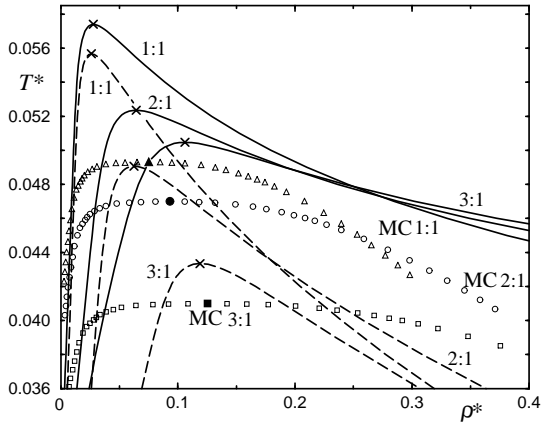}}
  \caption{\label{fig-diagphas}
    Coexistence curves computed for 1:1, 2:1, and 3:1
    equisize hard sphere ionic fluids or primitive model
    electrolytes: the solid lines correspond to the DHBjCI theory (without
    explicit hard-core excluded-volume terms);
    the dashed curves include `standard' bcc hard-core terms. The exclusion
    diameters used are the angular averages \eqref{angav} except
    for the refinement $a_4/a = 1.41$ for the 3:1 model. Solid
    symbols represent simulation estimates
    for the critical points \cite{young&mef04,pana&mef02} and open symbols
    the coexistence curves based on precise RPM simulations
    \cite{young&mef04}.}
\end{figure}

\begin{table}
\caption{\label{table-ys} Critical-point mole fractions, $y_\s$, of the 
  primary clusters (expressed as percentages) according to their total 
  charges, $q_\s$, for $z$:1 models described by DHBjCIHC theory (with 
  `standard' parameter values). Unlabeled clusters are monomers.}
\begin{ruledtabular}
\begin{tabular}{llllll}
 $q_\s/q_0=$ & $-1$ & $0$ & $+1$ & $+2$ & $+3$ \\
 1:1        & $9.14$ & $81.72$ & $9.14$ &---  & --- \\
            &        & (dimer) &  &  &  \\
 2:1        & $10.33$ & $72.93$ & $15.43$ & $1.31$ &  ---\\
            &        & (trimer) & (dimer) & & \\
 3:1        & $8.04$ & $77.17$ & $11.13$ & $3.32$ & $0.34$ \\
            &        & (tetramer) & (trimer) & (dimer) &
\end{tabular}
\end{ruledtabular}
\end{table}

Within this DHBjCIHC approach, we remark that an increase in 
the value of the
association constant $K_{1,1}$ yields a \textit{decrease} of $\Tce$
for a $1$:$1$ electrolyte: with the standard parameters, 
we find that varying $K_{1,1}$ by $\pm 5 \%$ around the value 
\eqref{k1z} results in changes of $\Tce(z \pe 1)$ of order $\mp 0.2\%$. 
While changing an association constant does not affect directly the
free energy of our model (indeed, $\Tce$ is the same in DH theories with or 
without ion association \cite{levi&mef96}), it nevertheless affects the 
various mole fractions, which do enter in the solvation free 
energies $\barf_\s^{\El}$. To our knowledge, the variation of the 
sum $\sum_\s \barf_\s^{\El}$ on increasing the association constant
can only be determined post-facto: for $z\pe 1$, our results indicate a 
more weakly coupled system with a lower critical temperature, in accord with
the findings of Jiang \textit{et al} 
\cite{jian&blum02}; however, for $z \pe 2$ and $3$, we find the opposite 
trend on varying $K_{z,z}$,  as noted below. 

Let us also recall that {\bI} presented numerical and graphical data showing 
how the predicted values of the critical parameters depend on the 
choices made for the mean ion separation $a_1$ and for the exclusion 
radius $a_2$. These results may reasonably be taken 
as indicative of the corresponding shifts that are likely to arise in 
our analysis of the 2:1 and 3:1 models.

\subsection{2:1 Hard Sphere Electrolyte}

We report first the basic DHBjCI results, using the angular averages $a_2$
and
$a_3$ listed in \eqref{angav}: they are
\begin{equation}
  \label{sq}
  \Tce(z \pe 2) =  0.05235, \quad \mbox{and} \quad 
  \roce(z \pe 2) =  0.06429 \, .
\end{equation}
The corresponding coexistence curve is plotted 
in Fig.~\ref{fig-diagphas}. One might note, first, that as in the 1:1
model, the shape of the liquid side of the coexistence curve 
below about $0.9 \, T_c (z \pe 2)$ becomes markedly concave. This behavior, 
while violating no known thermodynamic or other conditions, certainly 
appears unphysical. Furthermore, by comparison with the true results 
indicated by the simulations, this concavity must be judged as quite 
misleading. No doubt it results from the failure to satisfactorily describe 
the correlations, and thence, the free energy of the 
low-temperature liquid at densities $\rho^* \geq 0.15$ via a collection 
of free ions plus fairly compact neutral and singly charged clusters. 
This general issue is also addressed briefly in {\bI} Sec.~8.5: it may be 
noted 
that the standard MSA exhibits similar although somewhat less pronounced 
features: see {\bI} Fig.~8(d). 

On the other hand, the downward shift in $T_c$ and the marked increase 
in $\roc$ reproduce most satisfactorily both the trends and the 
magnitudes obtained in the simulations \cite{pana&mef02}: see Figs{.}
\ref{fig-Tc2z} and \ref{fig-rc2z}. 
These trends are also reproduced fully by the
other choices of exclusion diameters. However, as could be expected, 
the sensitivity to the size of the bigger clusters is enhanced in the 
2:1 case compared to the 1:1 model. Indeed, on using the steric diameters, 
we find $\Tce \pe  0.04850$ and $\roce \pe  0.0737$ implying a
drop by $7.3\%$ and an increase by $15\%$, respectively. With the harmonic 
parameters, the conclusions are reversed yielding $\Tce\pe 0.05574$ and 
$\roce \pe  0.06012$.  As discussed 
below (and see Table \ref{table}), a larger fraction of the
ions are bound in the clusters
when $z \pe 2$ compared to $z \pe 1$, thereby 
amplifying the sensitivity to the cluster characteristics. 

This enhanced sensitivity is also found for the 
hard-core effects: thus the values of $\Tce$ and $\roce$ predicted by the 
standard DHBjCIHC theory, listed in Table \ref{table}, are lower 
by $6\%$ and $3\%$ relative to the values in \eqref{sq}. Likewise,
the low density value of $B_\s$  yields 
$\Tce \pe  0.04375$ and $\roce \pe  0.06422$, which seriously 
overestimates the hard-core effects, making $\Tce$ drop  
to well below the Monte-Carlo estimate. However, 
the `optimal-fit'
choice $B_\s / \as^3 = 1.300$ yields the values
\begin{equation}
 \label{sc}
 \Tce (z \pe 2) = 0.04691 \quad \mbox{and}
 \quad \roce(z \pe 2) = 0.06285 \, ,
\end{equation}
which, indeed, provide the best
fit of our analysis to the Monte Carlo data (see
Figs.~\ref{fig-Tc2z} and \ref{fig-rc2z}). However, the corresponding
coexistence curve is excessively narrow even compared to the standard
prediction shown (dashed) in Fig.~\ref{fig-diagphas}.

On the other hand, it transpires that
the sensitivity to the association constant is not
so great. Thus in the DHBjCI approach, 
changing the cut-off for $K_{2,2}$ by $\pm 20\%$, induces changes in 
$K_{2,2}$ of order $\pm 1\%$, leading to shifts in 
$\Tce$ of order $\pm 0.003 \%$ and in 
$\roce$ of order $\pm 0.2\%$, totally negligible 
within our level of approximation. 

Returning to the standard DHBjCIHC theory, one sees from Table \ref{table} 
that it
predicts a drop in $\Tce$ (compared to the 1:1 electrolyte) of order
$12\%$ and an increase in $\roce$ of $140\%$. These 
results are to be compared with the Monte-Carlo results indicating 
a drop in $\Tce$ of $5\%$ and an increase in $\roce$ of around $24\%$.
The predicted $\Tce$ and $\roce$ agree within $4\%$ and $33\%$, 
respectively, with the current Monte Carlo estimates.   
The overall quantitative results are therefore fairly close to the Monte 
Carlo values, indicating that the main physical features have been 
captured by the theory. 

As regards the composition of the fluid at criticality, 
one learns from the last column of Table \ref{table} that fewer than 
$12\%$  of the ions now remain free or unassociated, even less 
than predicted in the 1:1 case. As can be seen from Table \ref{table-ys},
the free $+2 q_0$ ions are strongly depleted, less numerous than the 
$-q_0$ anions, by a factor $1/8$. Indeed, the numbers of positively charged 
dimers roughly match the oppositely charged free anions. However, while the 
predicted overall association rate is larger than for the RPM, the fraction 
of the ions bound into the neutral or ``molecular'' clusters 
(now trimers) is some $11\%$ smaller. 
Needless to say, the values of the $y_\s$ listed in Table \ref{table-ys}
verify the electroneutrality condition that 
implies $2 y_+ + \frac{1}{2} y_2 - y_- \pe  0$.


\subsection{3:1 Hard Sphere Electrolyte}

As before, let us first record the predictions of the basic DHBjCI 
theory using the angular averages, needed now for $a_2$, $a_3$, and 
$a_4$, the last for the tetramer which we expect to be the dominant 
species near criticality. We find
\begin{equation}
  \label{ss}
  \Tce (z \pe 3) = 0.05054 \quad \mbox{and} \quad \roce (z \pe 3) = 0.1063 
  \, ,
\end{equation}
where the corresponding coexistence curve is again displayed in 
Fig.~\ref{fig-diagphas}. These values, as is also clear from 
Figs.~\ref{fig-Tc2z}
and \ref{fig-rc2z}, continue to reproduce the appropriate trends with 
increasing $z$ as originally revealed by the Monte Carlo simulations. 
However, 
as also evident in Fig.~\ref{fig-Tc2z}, the drop in $\Tce$ of only 
$3.5\%$, relative to the 2:1 model, is significantly less than indicated by 
the simulations: in fact, the result \eqref{ss} suggests a concave variation 
for $\Tce (z)$ rather than the convex behavior maintained by the 
simulations up to $z \pe 4$ \cite{camp&pate99,pana&mef02}. 

The `culprit'
is obviously the failure to take explicit account of the hard-core
excluded volume effects important above and even \textit{at} the
predicted critical density which is $65\%$ larger than for the 2:1 model. 
The very slow decay of the liquid side of the coexistence curve when 
$\rho$ increases, as seen in Fig.~\ref{fig-diagphas}, strengthens the 
point. Indeed, the standard DHBjCIHC theory yields
\begin{equation}
  \label{sse}
  \Tce(z\pe 3) = 0.04580 \quad \mbox{and} \quad \roce(z\pe 3) 
   =  0.1089 \, . 
\end{equation}
Relative to the 2:1 model the critical temperature has now fallen by 
$12.5\%$, which may be compared with the Monte Carlo drop of $12.8\%$ 
(see Table \ref{table}). However, the value of $\roc$ has 
changed rather little. 

These predictions are not quite those entered in Table \ref{table} because 
it was deemed worthwhile for this case to explore further the influence of 
the exclusion radii. In particular, as discussed in {\bI} Sec.~6.3, the mean 
size of a physical cluster, for any sensible definition will grow  with 
increasing temperature. Hence, in choosing the tetramer exclusion radius 
$a_4$, it is reasonable to consider for use near $T_c$ a value somewhat
larger than the $T\pe 0$ angular average $a_4^a \pe 1.375 \, a$ 
[see \eqref{angav}]. Having examined the effects on the values of both 
$\Tce$ and $\roce$, the ratio 
\begin{equation}
  \label{sh}
  a_4 / a = 1.410 \,  ,
\end{equation}
was selected as a preferred refinement of the standard parameters. 
(The increase of $2.6\%$ brings the ratio to almost midway between $a_4^a/a$
and the steric value $a_4^s / a \! \simeq \! 1.452$.) Accordingly, 
\eqref{sh} has been adopted for computing the results displayed in
Table \ref{table}, in Figs.~\ref{fig-Tc2z} and \ref{fig-rc2z} and elsewhere
below; the corresponding coexistence curve for $z \pe 3$ is displayed
(dashed) in Fig.~\ref{fig-diagphas}. 

Evidently, the trends observed as $z$ increased from $1$ to $2$ are now 
continued regularly; and the previous concave variation of $T_c (z)$
is no longer so apparent. Furthermore, the trends still mirror rather 
faithfully those given by the simulations: These indicate a rise in 
$\roce$ by $35\%$ when $z$ changes from $2$ to $3$; the standard calculations
yield a $90\%$ relative rise which is significantly greater, but, as seen 
in Fig.~\ref{fig-rc2z}, not at all unreasonable. Indeed, $\Tce$ agrees
with the simulations to within $6\%$ while $\roce$ agrees to within 
$5\%$. Overall, both the magnitudes of $\Tce (z)$ and $\roce (z)$
and the trends with $z$ must be judged quite successful!

From Table \ref{table-ys} we see that the overall fraction of free 
ions remaining at criticality has now dropped still further to 
about $8.4\%$. At the same time close to three quarters of the ions are 
again bound in the neutral, molecular clusters (now tetramers). The 
fraction of free, unassociated cations of charge $+z q_0$ continues 
to fall dramatically as $z$ increases: on a heuristic basis, a 
decay like $y_{+,c} \! \sim \! e^{-bz}$ seems not implausible. Following 
the thought of Shelley and Patey \cite{shel&pate95}, one might also 
speculate that a system of rigid, neutral molecules or $(z\!+\!1)$-mers
formed of $z \! + \! 1$ equisize hard spheres with $z$ of charge $-q_0$
attached symmetrically to a central sphere of charge $+z q_0$, might 
continue 
to mimic the $z$:1 equisize hard-sphere ionic systems, at least up to 
$z \leq 12$. Beyond that, packing effects in the satellite ions could play 
an important role. 

It is probably appropriate to point out, as anticipated, that our
$z\pe 3$ predictions are less robust than those for $z \leq 2$. Thus the 
`optimal-fit' assignment $B_\s / \as^3 \pe 1.300$ together with the 
angular averages $a_2^a$ and $a_3^a$ but taking $a_4/a \pe 1.390$ yields
\begin{equation}
  \label{sn}
  \Tce (z \pe 3) = 0.04136 \quad \mbox{and} \quad
  \roce (z \pe 3) = 0.1171 \, ,
\end{equation}
which reproduces the Monte Carlo results quite satisfactorily. However, this 
choice once more leads to a coexistence curve which falls much too steeply
when $T< 0.9 \, T_c$. Again, the low-density value for $B_\s$ gives 
$\Tce < 0.038$ well below the simulation value. 

On the other hand, if one uses the $[5/2]$ Pad\'e approximant for the 
association constant integral $\I_{3,3}$ in \eqref{seriesItt} in place of the
more accurate fit \eqref{fitItt} one finds that the resulting 
$4\%$ decrease in $K_{3,3}$ [see Fig.~\ref{fig-pade}] leads to a decrease
in $\Tce$ of only $0.14\%$. The effect on $\roce$ is similar and hence, 
\textit{post facto}, of little consequence.

Finally, it is interesting to note from Table \ref{table} that the 
Debye length at criticality, namely, 
$\xi_{{\scriptscriptstyle{\rm{D}}}, c} \pe  1/ \kappa_c \pe  a / x_c$, 
decreases steadily as $z$ rises. In essence, this merely tells us that 
larger central charges in ionic clusters lead to tighter screening. 
It should be noted, however, that $\xid (T,\rho)$, as defined in 
\eqref{kappa} is not really susceptible to either physical measurement 
or simulation since our definition depends on having a well defined,
but intrinsically somewhat arbitrary decomposition of the system into 
distinct species of ionic clusters. On the other hand, the prediction 
that the critical pressure ratio, $Z_c \pe  p_c / \roc \kb T_c$, decreases 
strongly as $z$ increases (see Table \ref{table}, column 5) should be 
open to test by simulations.


\section{Special inflection loci} 
\label{chik}

In determining numerical values of critical parameters from a given model 
free energy, it is natural to start by calculating the two sides of the 
coexistence curve, $\rho_l (T)$ and $\rho_v (T)$, using techniques, such as 
illustrated in Fig.~\ref{fig-pmu}: in principle, one can then raise $T$ 
and monitor $\Delta \rho \! \equiv \! \rho_l (T) - \rho_v (T)$, determining 
$T_c$ from the vanishing of, say, $\Delta \rho^2$ and, then, $\rho_c$ from, 
say, $\frac{1}{2} (\rho_l - \rho_v)$ evaluated at $T_{c,{\rm{est}}}$. In 
practice, however, this method proves tedious and as experience (and a
consideration of 
Fig.~\ref{fig-pmu}) reveals is poorly adapted for providing precise and 
accurate (i.e., reliable!) values of $T_c$ and $\roc$.

An effective alternative is to confine attention to the one-phase region 
above $T_c$ where, in the first place, calculations are more straightforward
since, in particular, no `two-phase solutions' need be sought. Then, as 
demonstrated recently in simulations 
\cite{orko&mef01,young&mef&orko03,luij&mef02} (although also of value in 
studying experimental data) one may seek various loci, say $\rok(T)$, which 
all converge on the critical point. Since the isothermal compressibility 
$\chi_T = ( \partial \rho / \partial p )_T / \rho$ diverges at criticality, 
one obvious such locus is provided by those densities, say $\rho_0 (T)$, 
on which, at a fixed temperature above $T_c$, the compressibility achieves 
its maximum. But by considering the inflection points of the standard 
isothermal plots of $p$ vs volume or vs density, one soon realizes that this 
locus is but one of a natural family of $k$-loci, say $\rok (T)$, 
\cite{orko&mef01,young&mef&orko03,young&mef03}, on which the so-called 
$k$-susceptibilities, 
$\chik (T, \rho) \! \equiv \! \chi (T,\rho) / \rho^k$, attain their maxima: 
equivalently, these are just the loci of isothermal inflection points 
of plots of $p$ vs $\rho^k$. 

Because of their potential usefulness in simulation and experiment, the 
behavior of the $k$-loci in the scaling region close to criticality has been
investigated in some detail \cite{young&mef&orko03,young&mef03}. 
In the case of 
general, nonclassical critical points, they exhibit nontrivial and 
informative singular behavior as functions of $t\!\equiv\!(T-T_c)/T_c$ as 
$k$ varies. However, for classical critical behavior, as relevant here, 
\textit{all} the $k$-loci asymptotically approach the critical point 
$(T_c,\roc)$ \textit{linearly} in the $(T,\rho)$ plane. Thus by 
numerically determining two or three loci---for the results reported here
we utilized $k\pe 1$, $0$, and $-1$--- and solving for their mutual
intersection point, one may locate $T_c$ and $\roc$. In practice the method 
proves efficient and precise. 

More generally, however, the nature of the loci further from criticality and
a possibly characteristic dependence on $z$ is a matter of interest to which 
we now turn.


\subsection{Debye-H\"uckel predictions}

To gain a little perspective, let us examine, first, 
pure DH theory (as presented in Sec.~\ref{pure-DH}) where analytical 
calculations are feasible. Three cases arise as illustrated 
in Fig.~\ref{fig-chik-DH}. When $k \pe 1$, the pressure isotherm 
always has an inflection point above $T_c$, 
characterized by $\kappa a \pe x \pe 1$. The ($k\pe 1$) locus is thus 
a straight line starting at $(T_c, \roc)$, namely, 
$\rho^{(1)\, *} (T) \! \equiv \! T^*/4\pi$. When $k>1$ one sees that by 
construction, $\chik$ 
diverges to $+\infty$ when $\rho \rightarrow 0$ at fixed $T$; but 
this divergence competes with the localized
maximum driven by criticality. As a consequence, for $T$ above but not too
far from $T_c$, when $\rho$ drops beneath $\roc$ one first encounters
a maximum in $\chik$ and then a minimum before the divergence as
$\rho \rightarrow 0$. However, as $T$ is raised at fixed $k$ one
eventually encounters an annihilation or terminal point $(\Tak, \roak)$ at
which the minimum and maximum merge and the $k$-locus is terminated with
a horizontal slope, i.e. a tangent parallel to the $\rho$ axis. Above
$\Tak$ the susceptibility $\chik (T,\rho)$ falls monotonically as
$\rho$ increases and no `critical maxima' are realized.

\begin{figure}[t]
{\includegraphics[width=7cm,angle=-90]{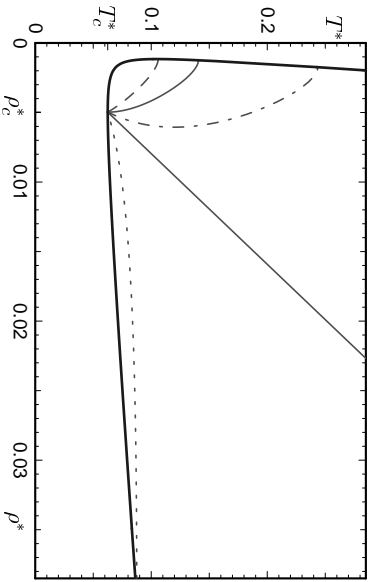}}
  \caption{\label{fig-chik-DH} Loci of the maxima of the
      $k$-susceptibilities in DH theory: all these $k$-loci intersect at
      the critical point, $\Tce \pe 0.0625$ and $\roce \! \simeq \! 0.00497$
      while the bold curve traces their termination points. The curved solid
      locus corresponding to
      $k=k_0=9/8$ displays a vertical slope at criticality while, the
      straight solid line corresponds to $k=1$. The dotted, dot-dash, and
      dashed curves are plots for $k \pe 0.70$, $1.06$, and $1.20$,
      respectively.}
\end{figure}

When $k<1$, a similar scenario emerges for $\rho > \roc$. As a result there
is, overall, a termination boundary in the $(\rho,T)$ plane with a
minimum at the critical point. For DH theory this takes the form of the
bold curve in Fig.~\ref{fig-chik-DH}. For $k<1$ the termination boundary
approaches asymptotically the line
$T_{a, k(<1)}^* \approx 4 \pi \rho^* / (2+\sqrt{3})^2$ while for $k>1$ and
large $\rho$ one has $T_{a, k(>1)}^* \approx 4 \pi \rho^* / (2-\sqrt{3})^2$.
As also clear from Fig.~\ref{fig-chik-DH} for values of $k$ differing much
from $1$, the $k$-loci are rather short (and hard to locate numerically).
The asymptotic slope of the general $k$-locus at criticality is given by
\begin{equation}
  \label{drmkdT}
  (d \rok / d T)_c = (2/\pi) (k_0 -k)\, ,
\end{equation}
where, within DH theory one has $k_0 \pe 9/8$ (for all $z$).


\subsection{DHBjCIHC Predictions}

How are these $k$-loci affected when the DH approximation is supplemented by
association, solvation and 
hard-core effects, and how do they evolve with $z$? Some results 
obtained with the full theory are displayed in Fig.~\ref{fig-rmk}. As 
expected from our analysis of 
the DH theory, most of the $k$-loci do indeed terminate with a 
horizontal slope at some point within the range of investigation. Most 
of the values of $k$ examined are smaller than $1$ and the 
$k$-loci bend towards high density. But, as in the pure DH theory [and 
as could be expected from the ideal high-$T$ low-$\rho$ limit, where 
$(\partial p / \partial \rho) \varpropto T$],
when $k>1$, the $k$-loci do indeed bend towards low densities. Furthermore, 
some loci are present only in a small neighborhood of the
critical point. Indeed the $(k\pe -1)$-locus
is not visible on the scale 
of Fig.~\ref{fig-rmk}, and the ($k\pe 0$)-loci are also quite small compared
to those for $k\pe 1$. In the 1:1 model treated without hard-core terms, 
the $k \pe 1$ and $k_0$ loci (with vertical tangent at criticality)
are quite extended, and the $k \pe 1$ locus extends to 
large values of $\rho$. However, these features are sensitive both to the
value of $z$ and the inclusion of the hard-core terms. Indeed, with
hard-core corrections the termination boundary rises rapidly when
$\rho>\roc$. Moreover,  when $z$  increases, the $k$-loci terminate sooner
when $T$ is raised. Likewise, the values of $k_0 (z)$, for which the
loci arrive vertically at the critical point depend strongly on $z$:
we find $k_0 \pe 0.93$, $0.18$, and $-0.87$, for $z \pe 1$, $2$, and $3$,
respectively. It would be interesting to test whether this trend is borne
out in simulations and whether it relates to the corresponding
Yang-Yang ratios \cite{yang&yang64,young&mef&orko03}.

\begin{figure}[t]
\scalebox{0.42}{\includegraphics{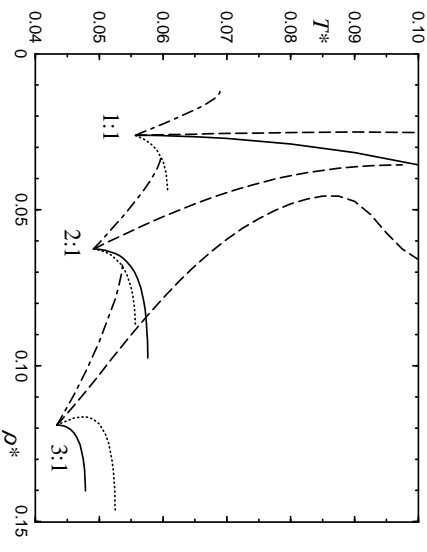}}
  \caption{\label{fig-rmk} Plots of the $k$-loci according to
    DHBjCIHC theory (using the
    preferred parameter values) for $z\pe 1$, $2$, and $3$.
    The solid lines correspond to the choice $k_0(z)$ yielding a vertical
    slope at criticality. The dotted, dashed, and dot-dash
    lines correspond to $k \pe 0$, $1$, and $2$, respectively.}
\end{figure}



\section{Interphase Galvani potential}
\label{galvani}

As mentioned in Sec.~\ref{phase-eq}, when charged species
are present in an equilibrium system, the \textit{electro}chemical
potentials  for each species must be equal in coexisting
phases. This necessitates the introduction of an overall, absolute
potential difference, $\Delta \phi$,
that must, in general, exist between distinct phases (even when only in 
possible rather than actual coexistence). The existence of such a potential, 
e.g., between an electrode and an electrolyte, is well recognized in the 
literature \cite{sparbock,iosi00,mura01,warr00} and is appropriately 
named a Galvani potential \cite{sparbock,iosi00}.
We have added the prefix ``interphase'' to
indicate that, speaking loosely, $\Delta \phi$ is spontaneously 
generated in an otherwise uniform medium when it decomposes into two 
(or more) phases beneath (or above) some critical point. 
Hence, equating the electrochemical potentials of both $+$ and 
$-$ species in the vapor and liquid phases gives
\begin{multline}
  \label{chem-eq2}
  \barmu_{+ v} + z \bphi_v=\barmu_{+ l} + z \bphi_l, \quad
      \barmu_{- v} - \bphi_v=\barmu_{- l} - \bphi_l  
      \\
 \quad  \mbox{with} \quad \bar{\phi}_\gamma \equiv \phi_\gamma q_0 / \kb T \, ,
\end{multline}
where $\phi_\gamma$ is
the electrostatic potential in phase $\gamma$. In fact, one soon
realizes that with the chemical equilibrium conditions (equating the
sum of the chemical potentials of the reactants and products), all these
equalities for the different species, are equivalent to any one of them.

The RPM is clearly a special case in which the gas-liquid interphase
Galvani potential vanishes identically for all $T$ owing to the
symmetry in charge and in all other interspecies interactions. 
As soon as this symmetry is broken in any way, the gas and liquid phases will
be distinguished by a non-zero Galvani potential.

To examine the issues further, let us adopt, first, the simplest
treatment, namely, pure Debye-H\"uckel theory with,
however, $z>1$ as  discussed in
Sec.~\ref{pure-DH}.   Using \eqref{mu-j} and
\eqref{f-dh}, and the electroneutrality $z \rho_+ \pe  \rho_-$, etc., the
partial chemical potentials for the positive and negative ions are,
respectively,
\begin{align}
  \barmu_+ = & -\frac{z x}{2 T^*(1+x)}+\ln (x^2 T^*)
  \nonumber \\ &
  +\ln \left( \frac{1}{1+z}
    \right) + \ln  \left( \frac{\Lambda_1^3}{4 \pi a^3} \right) \, ,\\
  \barmu_- = & -\frac{x}{2 z T^*(1+x)}+\ln (x^2 T^*)
  \nonumber \\ &
        + \ln \left(
 \frac{z}{1+z} \right) +\ln \left( \frac{\Lambda_1^3}{4 \pi a^3} \right)\, . 
\end{align}
By substituting in \eqref{chem-eq2} and
solving for the electrostatic potential difference, we obtain
\begin{equation}
  \label{EPD1}
     \Delta \bphi (T) = \frac{1}{z+1} \left [ z \ln \left( 
           \frac{\rho_{-l}}{\rho_{-v}} \right) 
     - \frac{1}{z}
        \ln \left(\frac{\rho_{+l}}{\rho_{+v}} \right) \right]\, ,
\end{equation}
where $\Delta \bphi \! \equiv \! q_0 \Delta \phi (T) / \kb T$ and 
$\Delta \phi = \phi_{liq} - \phi_{vap}$. 
On using the electroneutrality constraint, we obtain the much simpler form
\begin{equation}
  \label{EPD2}
  \Delta \bphi (T) = (1-z^{-1}) \ln[\rho_l(T) / \rho_v(T) ] \, .
\end{equation}
As anticipated, the predicted Galvani potential $\Delta \phi$ vanishes
identically when $z \pe 1$. 

Fig.~\ref{fig-galvani-DH} presents plots of this Debye-H\"uckel result 
for $\Delta \bphi$ vs. $T^*$ for various values of $z$. Note that when $T$ 
approaches $0$ the form \eqref{EPD2} implies that $\Delta \phi (T)$
should approach a constant value since $\rho_v (T)$ vanishes exponentially 
fast with $1/T$ \cite{mef&levi93}. We should remark that within DH theory
the ratio $\rho_l/\rho_v$ is independent of $z$ at fixed $T$. 
By expanding $\rho_l(T)$ and $\rho_v(T)$ around $\rho_c$ in
powers of $t \equiv (T_c-T)/T_c$, one finds 
$\Delta \phi \approx B_{\phi} t^{\beta}$, with since the theory 
is classical, $\beta \pe  \frac{1}{2}$.

\begin{figure}
\scalebox{0.43}{\includegraphics{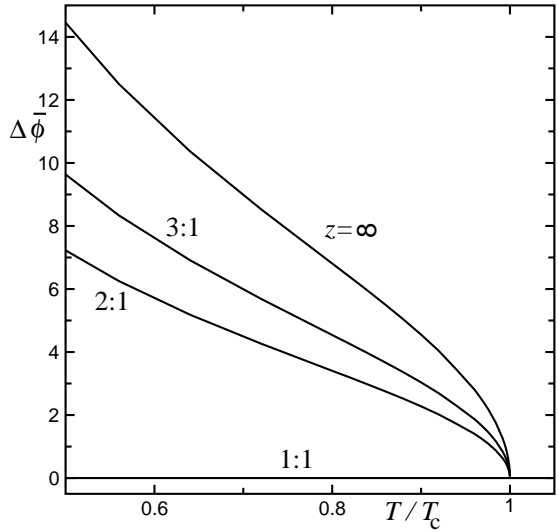}}
   \caption{\label{fig-galvani-DH} Plots of the reduced interphase Galvani
     potential $\Delta \bphi \pe q_0 \Delta \phi / \kb T$ for a $z$:1
     electrolyte as predicted by the pure Debye-H\"uckel theory.}
\end{figure}

\begin{figure}
\scalebox{0.43}{\includegraphics{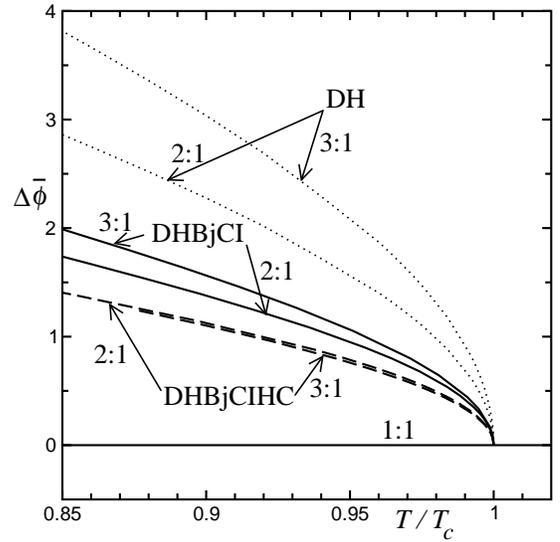}}
  \caption{\label{fig-galvani} The reduced interphase Galvani potential
    plotted vs $T/T_c$.
    The solid lines show the predictions of the DHBjCI theory
    (with $B_\s\pe0$), the dashed curves, the DHBjCIHC theory (with
    refined standard parameters), and the dotted plots, the DH theory.}
\end{figure}

In this simple DH analysis,
the Galvani potential is rather trivially proportional to the
logarithm of the ratio of densities in the two coexisting phases. One
might, perhaps, suspect that this indicates the existence of some simple
``universal'' result not depending significantly on the detailed
microscopic interactions. For a better understanding showing that this
idea is false, let us, following Bjerrum \cite{bjer26}, allow for the
formation
of dimers by association, neglect all solvation effects arising from
their dipole and higher moments, and treat the dimers as of the
same size as the free ions. This allows us to use the standard DH
free energy \eqref{fdh} for the electrostatic
contributions, and thence to write the total free energy density as
\begin{equation}
\barf(T^*;\{\rho_\s\})=\barf^{\nDH}(T^*;\{\rho_\s\})+\sum_{\s=+,-,2} 
\barf^{\Id}(\rho_\s)\, .
\end{equation}
Now the electrochemical equilibrium conditions always apply and thus,
Eqs.~\eqref{chem-eq2} give the Galvani potential correctly. Notice, however, 
that for $z >1$, the dimeric ion pairs carry a net charge $(z-1) q_0$ so 
that,
although electroneutrality must still be respected in both phases, the simple
ratio $\rho_+/\rho_-$, will, in general, be different in the liquid and 
the vapor. Consequently, the simple result \eqref{EPD2} no longer applies!
Clearly, the ratio of $\rho_+$ to $\rho_-$ depends on the density, 
$\rho_2(T)$,
of the dimers in the two phases. This, in turn, must depend via the 
mass-action laws, on the association constant $K_{1,z}(T)$ of the
dimers which then determines the overall degree of association, 
say, $\alpha_{2 \gamma} (T)$, which will vary very differently in each
phase $\gamma$. Accordingly, we may write $\rho_2 \pe  \alpha_2 (T) \rho$ 
and impose electroneutrality in both phases to 
simplify \eqref{EPD1}. The result for $\Delta \bphi$ may be written
\begin{widetext}
\begin{equation}
  \label{EPD3}
  \Delta \bphi= (1-z^{-1}) \ln \left( \frac{\rho_l}{\rho_v} \right)
  + \frac{z}{z+1} \ln \left(\frac{z-\alpha_{2l}(z+1)}{z-\alpha_{2v}(z+1)} 
       \right)
  - \frac{1}{z(z+1)} \ln \left(\frac{1-\alpha_{2l}(z+1)}{1-\alpha_{2v}(z+1)}
       \right)\, ,
\end{equation}
\end{widetext}
which, by comparison, demonstrates that, in general, the simple 
form \eqref{EPD2} must be modified by nontrivial temperature-dependent 
terms that depend on the details of the ionic interaction, etc. Nevertheless,
the predicted leading temperature variation will still reflect the $t^\beta$ 
form characterizing the coexistence curve. The analysis 
leading to \eqref{EPD3} involved only the formation of non-neutral 
dimers; but it is clear that in any realistic treatment there will be a
variety of charged species present in temperature-varying 
proportions determined by microscopic details. Thus simple results for the 
interphase Galvani potential should not be anticipated. 

On the other hand, from our explicit calculations of the coexistence curves 
for the 2:1 and 3:1 models, we may determine $\Delta \phi(T)$ via 
\eqref{chem-eq2}, merely by computing the
difference in $\mu_-$ in the vapor and liquid phases (which 
quantity arises naturally in the computations). The results are
presented in Fig.~\ref{fig-galvani}, where the temperatures 
have been normalized by the respective critical temperatures to
facilitate comparison. The plots are qualitatively similar to those 
predicted by the pure DH theory. However, we note that in the full theory 
with standard parameters, it is not possible to draw a conclusion 
regarding the trend of $\Delta \bphi$ with $z$.

The observability of $\Delta \phi(T)$ in a real system is
elusive if not in principle impossible \cite{sparbock,iosi00}; 
however, it seems 
that it should be possible to measure  $\Delta \phi(T)$ in 
simulations. Specifically, the potential distribution theorem of Widom
\cite{wido63,rowl&wido89}
provides a direct way of sampling the (absolute)
electrochemical potential via a  suitably weighted 
average interaction of a ``ghost test
particle'' with the interacting ions in the system which do not ``see'' the 
ghost particle. The
electrochemical potential of a (ghost) ion of specific charge should
thereby be open to estimation in liquid-like and vapor-like
simulations of the restricted primitive models (or more general
models). The appropriate difference should then provide a value of
$\Delta \phi(T)$.



\section{Discussion}
\label{discussion}

Our aim has been to understand, both qualitatively and semi-quantitatively,
the role of charge asymmetry in the criticality of electrolytes. We have
extended the DHBjDIHC theory of Fisher and Levin
\cite{mef&levi93,levi&mef96} for 1:1 electrolytes to 2:1
and 3:1 electrolytes by accounting for association of ions into charged
clusters and including the interaction  of the clusters with the screening 
ions (solvation). Thus we have labeled the extended theory: 
DHBj\textbf{CI}HC, where the CI now stands for the cluster-ion interactions
and, for a $z$:1 system, explicit account has been taken of the monomers, 
with charges $-q_0$ and $+z q_0$, of dimers, trimers, $\ldots$, up to 
neutral $(z\!+\!1)$-mers. The principal results, summarized in 
Figs.~\ref{fig-Tc2z} and \ref{fig-rc2z},
indicate that the reduced critical temperature, $\Tce (z)$, 
\textit{decreases} while the critical
density \textit{increases} with increasing charge asymmetry. 
Furthermore, these trends and the magnitudes of the changes with $z$ 
agree with the behavior revealed by computer simulations and present
a significant improvement over the original DH theory.

To understand the results in physical terms, consider, first,
the pure DH theory which
predicts that the critical temperature and density are {\em
independent} of charge asymmetry: as shown in Figs.~\ref{fig-Tc2z} and 
\ref{fig-rc2z}. The only direct attractive interactions accounted for 
in this theory are those between the ions of opposite charge. These induce 
a Debye screening cloud around each (monomeric) ion and the associated 
`solvation free energy' drives the vapor-liquid phase separation below 
$T_c (z)$, the vapor phase being stabilized by the greater entropy available 
at low densities. 

The temperature is appropriately normalized by the energy of attraction
of the opposite ions at contact, namely,  
$\varepsilon \pe |q_+ q_-|/Da$. Under this normalization [see \eqref{T*}] the
maximum strength of the attractive interactions is always $\varepsilon$ and
the pure DH theory therefore predicts that the (reduced) critical
temperature, $\Tce (z)$,  is independent of $z$. 

The DHBjCIHC theory, however, also takes
into account the formation of ion clusters and treats them as
distinct species, albeit in mutual chemical equilibrium which calls for the 
calculation of association constants.  
For example, for 2:1 electrolytes, the dimers
are species with charge $+q_0$ while the trimers are neutral. The two 
principal attractive interactions in this case are of magnitude
$\varepsilon$ between the positive and negative free ions, but only 
$\simeq \frac{1}{2} \varepsilon$ between the dimer and the negative ion.  
(The interaction magnitude is not precisely $\frac{1}{2} \varepsilon$ since the
dimer has a different effective exclusion zone radius, i.e., 
$a_2 \! \ne \! a$.)
Thus, relative to the 1:1 case, the effective attractions
are smaller for 2:1 electrolytes which explains why the critical
temperature should be expected to 
decrease. The same argument applies for larger $z$ when 
there are more intermediate positively
charged species between the free positive ions and the neutral
clusters. The strongest interactions is between two free, oppositely charged 
monomers and is always of magnitude 
$\varepsilon$: thus the overall effective interaction
decreases with increasing $z$ and the critical temperature decreases
correspondingly.

To understand the trend exhibited 
by the critical density, $\roce (z) \pe \rho_c a^3$, one must focus on the 
role played by the neutral clusters. 
As originally shown by Fisher and Levin for 1:1 electrolytes,
the association of free ions into neutral dimers is highly  significant
at criticality. Indeed, according to our theoretical estimates they constitute
about $82\%$ of the overall ion density. 
For  2:1  electrolytes, our analysis likewise indicates that about 
$73\%$ are bound in neutral trimers while for 3:1 systems the figure is 
$77\%$ for the neutral tetramers. As a consequence, not only are the 
relative effective charges of the charged species decreased by association 
(as just argued) but, in addition, the overall effective 
fraction of ions in charged
clusters is diminished when $z$ increases. In leading approximation the
solvation of a given cluster (charged or neutral) is achieved only by 
charged species. To obtain comparable solvation free energy
therefore necessitates higher overall ion densities and, thereby, an 
associated increase in critical density. 
The effect is reflected more concretely in the expression \eqref{kappa} 
for the effective inverse Debye length, $\kappa (T,\{\rho_\s\})$, whose 
critical value similarly increases with $z$: see Table \ref{table}.

This accounts for the trends displayed in Fig.~\ref{fig-rc2z}. It is 
interesting to notice, however, that while the Monte Carlo results 
display similar increases in $\roce (z)$, the magnitudes of the increases are 
rather smaller. It seems likely that this is associated with our neglect 
of the solvating influence of the neutral clusters which may be envisaged as
contributing to a change in effective dielectric constant. However, the 
increasing sensitivity of the results to the explicit 
hard-core excluded-volume terms when $z$ increases must also be noted. 

It is also appropriate to recall here that our present analysis takes no 
account  of critical fluctuations. Extensive studies demonstrate that the
effect of the fluctuations is to lower $T_c$ by $5$-$10\%$ or more 
relative to basic mean-field-type theories while having little effect on 
the slope of the coexistence curve diameter. In addition, the coexistence curve
is flattened (since $\beta \! < \! \frac{1}{2}$). As evident from 
Fig.~\ref{fig-diagphas} the present calculations are quite consistent with
these general expectations. 

Our results for $T_c (z)$ and $\rho_c (z)$ 
 are contrasted with those of other available theories in 
Figs.~\ref{fig-Tc2x} and \ref{fig-rc2x}. For this comparison, 
we have used the standard parameters of the  
DHBjCIHC theory (with a refinement for $z \pe 3$) as described in Table 
\ref{table} and Sec.~\ref{results}. We may note, first, that the mean 
spherical approximation (MSA) \cite{sabi&bhui98,gonz99}, like the original DH
theory \cite{deby&huck23}, predicts that $T_c^*$ and $\rho_c^*$ remain {\em
independent} of $z$. This seems primarily due to the failure to take ion
association in a sufficiently explicit way.  A field-theoretic expansion 
approach advanced by Netz and Orland (NO, dashed curves) \cite{netz&orla99}, 
in which the particle hard-cores are represented by a sharp, large-wavelength
cut-off, predicts that $T_c^*(z)$ \textit{increases} strongly with $z$
while $\rho_c^*(z)$ falls precipitously at small $z$ ($<1$) 
and then rises slowly.
In fact, the only previous theory known to us that matches the sign of the
trends
revealed by the simulations is the symmetric Poisson-Boltzmann (SPB, crosses)
integral equation analyses by Sabir, Bhuiyan, and Outhwaite 
\cite{sabi&bhui98}. However, not only is the critical temperature for the RPM
predicted by the SPB theory 
significantly too high (at $T_c^* \pe 0.0715$) but the
proportionate changes with $z$ are quantitatively much too small (by
factors of $5.6$ and $6.4$ for the 2:1 and 3:1 models,
respectively). Furthermore, the modified  Poisson-Boltzmann (MPB)
approach developed by the same authors, which they argue should be
quantitatively
and qualitatively better than the SPB, predicts the opposite trend 
for $\Tce (z)$. 
Finally, we note that recently devised  mean field theories based on 
Kac-Siegert-Stratonovich-Hubbard-Edwards transformations of the Boltzmann
factor \cite{cail05}, lead to critical temperatures which increase 
significantly with charge asymmetry again in strong contradiction to the 
Monte Carlo estimates (open circles in Figs.~\ref{fig-Tc2x} and 
\ref{fig-rc2x}).

\begin{figure}
\scalebox{0.43}{\includegraphics{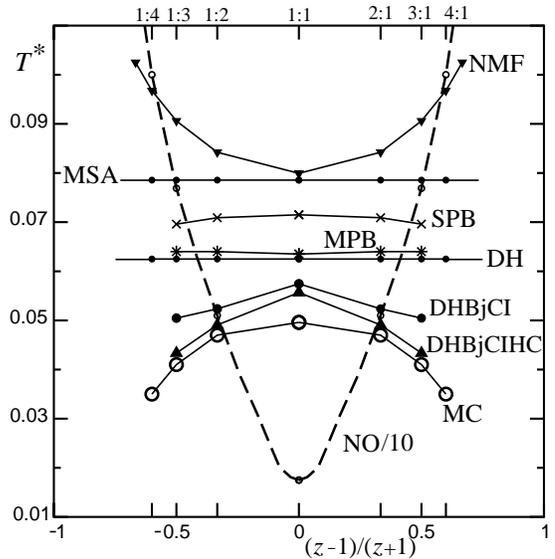}}
  \caption{\label{fig-Tc2x}
    Reduced critical temperature, $\Tce (z)$, as a function of the charge
    asymmetry parameter $w \pe (z-1)/(z+1)$ as found by simulations
    (open circles) \cite{camp&pate99,pana&mef02} compared with the present
    calculations (filled circles and triangles) and other current
    theories: MSA, SPB, and MPB \cite{sabi&bhui98,gonz99}, a
    `new mean-field theory' (NMF) \cite{cail05}, and a field theoretic
    expansion (NO, dashed curve) \cite{netz&orla99}; note that the
    predictions of the field theoretic approach
    have been divided by $10$ to bring them within the compass of the figure.}
\end{figure}

\begin{figure}
\scalebox{0.43}{\includegraphics{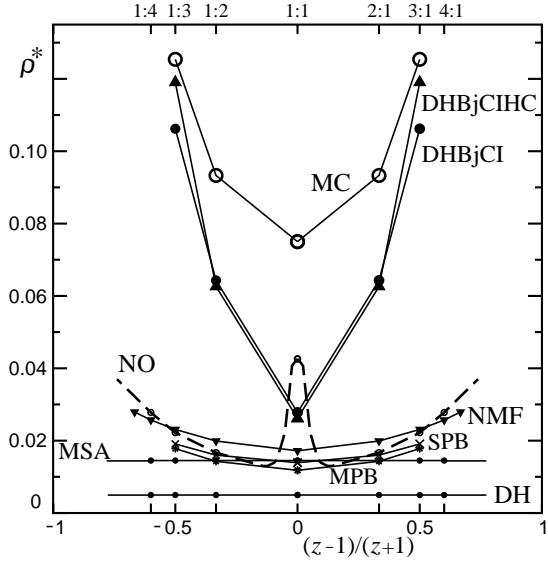}}
  \caption{\label{fig-rc2x}
    Simulation estimates (open circles) for the critical density, $\roce(z)$,
    compared with those of the present calculations (filled circles) and of
    other approaches: the labels, symbols, etc., have the same significance
    as in Fig.~\ref{fig-Tc2x}.}
\end{figure}

While our theoretical analyses have been
based upon fundamental principles and
provide insight into the variation of the critical parameters of
charge-asymmetric primitive model
electrolytes, it must be recognized that the results rest upon various
approximations. Thus, one of
our main approximations entails the choice of an equivalent sphere to
represent the
exclusion domain of a cluster. Moreover, we have not explicitly
considered higher order
association. That, despite these and other approximations, we find both the
correct trends and reasonable
quantitative agreements with the Monte Carlo simulations,
reinforces our conclusion that the main physical features linked to
charge asymmetry have been appropriately
captured by the theory.

\begin{acknowledgments}
The interest of Thanos Panagiotopoulos and Yan Levin has been appreciated. 
We are grateful to Patrick B. Warren and Joel L. Lebowitz for discussions 
pertaining to the Galvani potential and to Mikhail A. Anisimov for 
bringing the work of Muratov \cite{mura01}
to our attention. We are indebted  to Young C. Kim for 
discussions and for assistance with the Monte Carlo simulation results. 
We also thank Paul Sinclair, of Rhodes College, for his collaboration 
concerning the
Monte Carlo calculations of $\I_{3,3}$, together with the 
Rhodes College Information and Technology department who made their
computers available at short notice. 
The support of the National Science Foundation 
[under Grants No. CHE 99-81772 and 03-01101] is gratefully acknowledged.   
\end{acknowledgments}

\appendix



\section{Association constant for the tetramer}
\label{appa}

Consider a cation with charge $q_+ \pe z q_0$ at the origin. Without loss
of generality let the first satellite charge $-q_0$ be on the $x$-axis
at $\vecr_1 \pe (r_1, 0, 0)$ in Cartesian coordinates as shown in
Fig.~\ref{fig-tetramer}. Taking advantage of the azimuthal
symmetry, let the second satellite charge be in the $x$-$y$ plane
at $\vecr_2 \pe  (r_2 \cos \theta_{12}, r_2 \sin \theta_{12}, 0)$, where
$\theta_{12}$ is the angle subtended by the satellite pair $(1,2)$
at the origin.
Then the most general coordinates for the third satellite are
$\vecr_3 \pe  (r_3 \cos \theta_{13}, -r_3 \sin \theta_{13} \cos \varphi,
-r_3 \sin \theta_{13} \sin \varphi )$, where
$\theta_{13}$ is the angle subtended by the satellite pair $(1,3)$ at the
origin and $\varphi$ is the angle between the $(1,2)$ and $(1,3)$ planes
with
$\varphi \pe 0$ representing the planar configuration.

The ground state is clearly given by $r_1 \pe r_2 \pe r_3 \pe a$,
$ \theta_{12} \pe \theta_{13} \pe 2\pi/3$,
and $\varphi \pe 0$. Noting that the main contribution to the integral
defining $K_{3,3}$ [see \eqref{assoc-const}] comes
from near $T \pe 0$, it is helpful to  define rescaled coordinates
\begin{equation}
  \begin{split}%
  \theta_2^* \equiv &%
   (\theta_{12}-2 \pi/3) / \sqrt {z T^*} \, ,
  \\
  \theta_3^*  \equiv &%
   (\theta_{13}- 2 \pi/3) / \sqrt {z T^*}\, ,
  \\
  l_i \equiv &%
   (r_i/a-1) / \Te \, ,
  \end{split}%
\end{equation}
for $i\pe 1, 2, 3$. Then, by expanding about the ground state configuration
for small $\Te$, one can write the configurational energy to leading order as
\begin{multline}%
  \label{app2}
       \frac{E_{3,z}}{\Te} =
          \frac{3 C_{3,z}}{\Te} - C_{3,z} \sum_{i=1}^3 l_i
        - \frac{1}{8\sqrt 3}{\varphi^*}^2
        \\%
        - \frac{5}{12\sqrt{3}} ({\theta_1^*}^2 +  {\theta_2^*}^2
           + {\theta_1^*} {\theta_2^*})
         + \cO (\sqrt{\Te})\, ,
\end{multline}%
where $C_{3,z} \pe 1-1/\sqrt{3} \, z$. The infinitesimal phase-space
volume can likewise be written
\begin{multline}%
  \label{app3}
  d\vecr_1 d\vecr_2 d\vecr_3 = a^9 {\Te}^3 dl_1 dl_2 dl_3
  \\
  \times 8 \pi^2
  \sin^2 (2 \pi/3) (z T^*)^{3/2} d{\theta_1^*} d{\theta_2^*} d{\varphi^*}
  [1+\cO(\Te)]\, .
\end{multline}%
To evaluate the defining integral in \eqref{assoc-const} we diagonalize the
angular quadratic form in \eqref{app2} by introducing coordinates
\begin{equation}
  \label{app4}
  X\pe (\theta_2^*+\theta_3^*)/{\sqrt 2} \quad \mbox{and} \quad  
  Y\pe (\theta_2^*-\theta_3^*)/{\sqrt 2} \, ,
\end{equation}
 to obtain 
\begin{equation}
  \label{app}
  {\theta_1^*}^2 + {\theta_2^*}^2 + {\theta_1^*} {\theta_2^*} =
  \mbox{$\frac{1}{2}$} (3 X^2 + Y^2) \, .
\end{equation}         
The integrals in~\eqref{assoc-const} can then be evaluated in the
form \eqref{kmz}, with Jacobean and eigenvalues 
\begin{equation}
  \mathcal{J}_3 = \mbox{$\frac{3}{4}$}, \quad {\rm and } \quad \{ 
  \lambda_{3,k}
     \} =    
  \left \{ \frac{1}{8{\sqrt 3}}, \frac{5}{24\sqrt{3}}, \frac{5}{8\sqrt{3}} 
     \right \}\, .
\end{equation}       
Interestingly, the $\sqrt{\Te}$ corrections (and all subsequent half-integer
power-law corrections) arising in \eqref{app2} vanish upon integration
because they are all associated with odd powers of the angular variables. 
An expansion for $\I_{3,z}$ can then be found by carrying
the expansions in \eqref{app2} and \eqref{app3} to higher order in $T^*$, 
and, hence, to higher orders in the $l_i$ and in $\varphi$, $\theta_1$
and $\theta_2$. The resulting Gaussian integrals can be
performed --- analytically in low orders and numerically, with increasing 
difficulty, in the higher orders --- leading to the asymptotic 
expansion \eqref{seriesItt}.


\section{Monte Carlo evaluation of the tetramer association constant}
\label{MC}

To evaluate $K_{3,3}$ numerically, in order to validate the Pad\'e 
approximants constructed from \eqref{seriesItt} and to correct them 
at higher temperatures, we undertook Monte Carlo integration computations 
following accepted procedures \cite{goul&tobo}. However,  
the general sample-mean method, at first yields results with errors 
significantly too large at the small values of $\Te$ needed 
for ionic criticality. The reason is simply that the integrand of $K_{3,3}$
is sharply peaked around the ground state, the peak sharpening as
$\Te$ is lowered and hence becoming less frequently sampled. To improve 
the accuracy, we used a `weighted sample-mean
method', in which random numbers are generated with a weighting chosen to  
sample the integrand more often near the peak. Thus, in one 
dimension, for example, to evaluate $I \pe  \int_a^b f(x) dx$, one needs to 
calculate 
\begin{equation}
  \label{}
  I_n = \frac{1}{n} \sum_{i=1}^n \frac{f(x_i)}{p(x_i)}\, ,
\end{equation}
where $p(x)$ is the probability density function used to generate the
random numbers, normalized in the interval $[a,b]$, and $n$ is the total 
number of random points $x_i \in [a,b]$. The weighted random numbers 
are generated from the uniform random numbers $\sigma \in [0,1]$ by 
solving for $x$ in 
\begin{equation}
  \label{coucou}
  P (x) \equiv \int_{-\infty}^x p(x') \, dx' = \sigma \, . 
\end{equation}
The density function should be chosen so that this relation 
can be solved for $x$ algebraically. 

We generalized this procedure to the geometry of the tetramer. For the 
radial integrals, we used the density function 
\begin{equation}
  \label{}
  p(r) = A_r \exp ( - \lambda_r \, r) \quad \mbox{with} \quad 
  \lambda_r \pe  C_{3,3}/\Te \, , 
\end{equation}
and $A_r \pe  1/\int_a^R \exp(-\lambda_r \, r) \, dr$. 
This weighting mimics the peaks in the integrand almost exactly.
For the angular variables the optimal 
weighting is more complicated because the 
peaks are Gaussian leading to an equation \eqref{coucou} that cannot be 
simply inverted algebraically. Instead, we used exponential weighting 
\begin{equation}
  \label{B4}
  p(\omega) = A_\omega \exp (- \lambda_\omega |\omega|)\, ,
\end{equation}
with $\omega \pe  \varphi$, $\theta_2 \pe  \theta_{12}\! - \! 2\pi/3$ or 
$\theta_3 \pe  \theta_{13} \!- \!2\pi/3$, and with the normalizing 
integrals $A_\varphi^{-1} \pe  \int_0^\pi \exp(- \lambda_\varphi |\varphi|)$ 
and 
$A_\theta^{-1} \pe \int_{-2\pi/3}^{\pi/3} \exp (-\lambda_\theta |\theta|)$.
We also chose $\lambda_\varphi \pe  1/(24 \sqrt{3} \Te)^{1/2}$ and 
$\lambda_\theta \pe  2.5 /(24 \sqrt{3} \Te)^{1/2}$ so that the width
of the peak in \eqref{B4} matched the width of the peak of the integrand.
Finally, as a generalization of the Bjerrum procedure,
we used a radial cut-off $R \pe  0.196 \, a/\Te$ which satisfactorily
located the minimum of $\partial K_{3,3} / \partial R$.

The results, which are well fit by \eqref{fitItt} with the coefficients
listed in Table~\ref{tableij}, agree closely with the consensus of the
seventh order Pad\'e approximants up to $T^* \simeq 0.03$; but they deviate
strongly above $T^* \pe 0.06$: see Fig.~\ref{fig-pade}.


\begin{thebibliography}{50}
\bibitem[{\citenamefont{Camp and Patey}(1999)}]{camp&pate99}
\bibinfo{author}{\bibfnamefont{P.~J.} \bibnamefont{Camp}} \bibnamefont{and}
  \bibinfo{author}{\bibfnamefont{G.~N.} \bibnamefont{Patey}},
  \bibinfo{journal}{J. Chem. Phys.} \textbf{\bibinfo{volume}{111}},
  \bibinfo{pages}{9000} (\bibinfo{year}{1999}); \bibinfo{note}{note that we
  cite this study here because, to our knowledge, it provides the only
  available estimate for $T_c^*$ (but not for $\rho_c^*$) for the case
  $z\!=\!4$.}
\end{thebibliography}



\end{document}